# One-year reserve risk including a tail factor : closed formula and bootstrap approaches

Alexandre Boumezoued
R&D Consultant – Milliman Paris
alexandre.boumezoued@milliman.com

Yoboua Angoua
Non-Life Consultant – Milliman Paris
yoboua.angoua@milliman.com

Laurent Devineau
Université de Lyon, Université Lyon 1, Laboratoire de Science Actuarielle et Financière,
ISFA, 50 avenue Tony Garnier, F-69007 Lyon
laurent.devineau@isfaserveur.univ-lyon1.fr
Head of R&D – Milliman Paris
laurent.devineau@milliman.com

Jean-Philippe Boisseau
Non-Life Senior Consultant – Milliman Paris
jean-philippe.boisseau@milliman.com

ABSTRACT

In this paper, we detail the main simulation methods used in practice to measure one-year reserve risk, and describe the bootstrap method providing an empirical distribution of the Claims Development Result (CDR) whose variance is identical to the closed-form expression of the prediction error proposed by Wüthrich *et al.* (2008). In particular, we integrate the stochastic modeling of a tail factor in the bootstrap procedure. We demonstrate the equivalence with existing analytical results and develop closed-form expressions for the error of prediction including a tail factor. A numerical example is given at the end of this study.

# Contents

# 1. Introduction

Solvency II is the updated set of regulatory requirements for insurance firms which operate in the European Union. It is scheduled to come into effect late 2012. Solvency II introduces strong changes comparing to prudential rules currently in force in Europe (Solvency I). These new solvency requirements will be more risk-sensitive and more sophisticated than in the past. Thus, the Solvency Capital Requirement (SCR) shall correspond to the Value-at-Risk of the basic own funds of an insurance or reinsurance undertaking subject to a confidence level of 99.5 % over a one-year period. The Solvency Capital Requirement shall cover at least several risks, including non-life underwriting risk. The non-life underwriting risk module shall reflect the risk arising from non-life insurance obligations, in relation to the perils covered and the processes used in the conduct of business.

One of the main sources of uncertainty for a non-life company is the estimation of its insurance liabilities, in particular the amount of claims reserves. In the Solvency II framework, claims reserves have to be evaluated based on a "best estimate" approach. The best estimate should correspond to the probability weighted average of future cash-flows taking account of the time value of money. The uncertainty regarding this evaluation essentially arises from the fact that the amount of future payments relative to incurred claims is unknown at the valuation date. We focus in this study on the measure of this uncertainty (reserve risk).

Solvency capital calculations will be based on the standard formula or an internal model. When using an internal model, risks calibration (in particular the calibration of reserve risk) has to rely on internal data and insurance companies have to define their own methodology to measure risks. Regarding the evaluation of the reserve risk, undertaking-specific parameters may also be used. In order to use an internal model or undertaking-specific parameters, insurance companies are required to ask for supervisory approval first.

Given the time horizon defined in the Directive, one of the major issues raised by Solvency II to non-life insurance undertakings is to understand how to measure volatility in their claims reserves over a one-year time horizon.

Insurance undertakings generally use stochastic reserving methods that enable them to measure volatility in their "best estimate" evaluation. The two most common methods are the model of Mack (1993) and the bootstrap procedure. However, these methods provide in their "standard" version an ultimate view of the claims reserves volatility and not a one-year view as required per the Solvency II framework.

The model of Wüthrich *et al.* (2008) is the first model to meet the one year time horizon in order to measure the volatility in claims reserves. This method, giving a closed-form expression of the one-year volatility of claims reserves, is one of the standardized methods for undertaking-specific parameters for reserve risk.

The goal of this study is to assess an alternative methodology to the model of Wüthrich *et al.* (2008) to measure the uncertainty of claims reserves over a one-year time horizon. This alternative method is based on the bootstrap procedure.

Compared to the model of Wüthrich *et al.* (2008), the bootstrap adaptation outlined in this study has many advantages:
- The model replicates the results of Wüthrich *et al.* (2008),
- When using an internal model, the method allows to obtain a distribution of one-year future payments and a distribution of the best estimate of claims reserves at time $t = I + 1$,
- The method also allows to take into account a tail factor, including a volatility measure with regard to this tail factor.

## 2. Reserve risk: definition and measure

Reserve risk corresponds to the risk that technical provisions set up for claims already occurred at the valuation date will be insufficient to cover these claims. A one-year period is used as a basis, so that the reserve risk is only the risk of the technical provisions (in the Solvency II balance sheet) for existing claims needing to be increased within a twelve-month period.

Let's consider an insurance company which faces the reserve risk only. By simplification, we will ignore the discount effect, and we will not take into account risk margin and differed tax.
We define the "best estimate" as the probability weighted average of future cash-flows, without prudential margin, and without discount effect.

Thus, the NAV (*Net Asset Value*) is given by:
$$NAV(I) = A(I) - \hat{R}_i^I = A(I) - (\hat{C}_{i,I_{ult}}^I - C_{i,I-i}),$$
and
$$NAV(I+1) = A(I+1) - \hat{R}_i^{I+1} = A(I) - X_{i,I-i+1} - (\hat{C}_{i,I_{ult}}^{I+1} - C_{i,I-i} - X_{i,I-i+1}),$$
with

- $A(I)$: market value of assets at $t = I$,

- $NAV(I)$: Net Asset Value at $t = I$,

- $NAV(I+1)$: Net Asset Value at $t = I + 1$,

- $\hat{C}_{i,I_{ult}}^I$: best estimate of the total ultimate claim for accident year $i$, given the available information up to time $t = I$,

- $\hat{R}_i^I$: best estimate of claims reserves for accident year $i$, given the available information up to time $t = I$,

- $\hat{C}_{i,I_{ult}}^{I+1}$: best estimate of the total ultimate claim for accident year $i$, given the available information up to time $t = I + 1$,

- $\hat{R}_i^{I+1}$: best estimate of claims reserves for accident year $i$, given the available information up to time $t = I + 1$,

- $X_{i,I-i+1}$: incremental payments between $t = I$ and $t = I + 1$ for accident year $i$,

- $C_{i,I-i}$: cumulative payments at $t = I$ for accident year $i$.

The insurance company only faces to the reserve risk. Thus, we have:

$$SCR_{reserve} = NAV(I) - VaR_{0,5\%}(NAV(I+1)),$$
$$SCR_{reserve} = A(I) - (\hat{C}^I_{i,I_{ult}} - C_{i,I-i}) - VaR_{0,5\%}\left(A(I) - X_{i,I-i+1} - (\hat{C}^{I+1}_{i,I_{ult}} - C_{i,I-i} - X_{i,I-i+1})\right),$$
$$SCR_{reserve} = A(I) - \hat{C}^I_{i,I_{ult}} + C_{i,I-i} - A(I) - C_{i,I-i} - VaR_{0,5\%}(-\hat{C}^{I+1}_{i,I_{ult}}),$$
$$SCR_{reserve} = -VaR_{0,5\%}(\hat{C}^I_{i,I_{ult}} - \hat{C}^{I+1}_{i,I_{ult}}),$$
$$SCR_{reserve} = -VaR_{0,5\%}\left(\hat{R}^I_i - (\hat{R}^{I+1}_i + X_{i,I-i+1})\right).$$

The amount $\hat{C}^I_{i,I_{ult}} - \hat{C}^{I+1}_{i,I_{ult}} = \hat{R}^I_i - (\hat{R}^{I+1}_i + X_{i,I-i+1})$ corresponds to $\widehat{CDR}_i(I+1)$. The Claims Development Result (CDR) is then defined to be the difference between two successive predictions of the total ultimate claim. This definition has been introduced for the first time by Wüthrich *et al.* (2008).

The changes in the economic balance sheet between $t = I$ and $t = I + 1$ are shown in Figure 1.

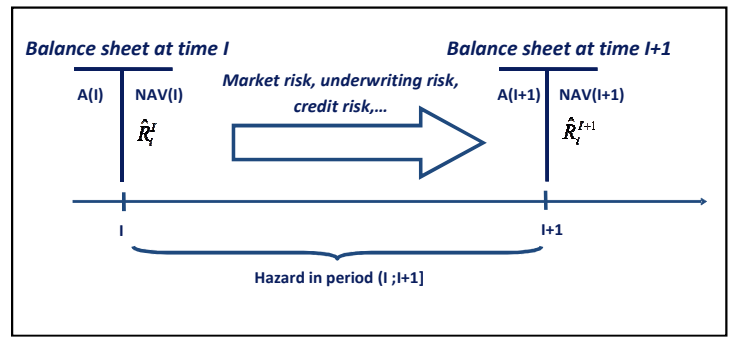

**Figure 1:** Economic balance sheet over a one-year period.

Thus, the Solvency Capital Requirement for the reserve risk is equal to the opposite of the 0.5%-percentile of the CDR distribution.
Solvency capital calculations will be based on the standard formula or an internal model. Regarding the evaluation of the reserve risk, undertaking-specific parameters may also be used.

**Standard formula**
In the standard formula framework, the reserve risk is part of the Non-life premium & reserve risk module. The evaluation of the capital requirement for the Non-life premium & reserve risk module is based on a volume measure and a function of the standard deviations given for each line of business. For reserve risk, the volume measure corresponds to the best estimate of claims reserves. This amount should be net of the amount recoverable from reinsurance and special purpose vehicles, and should include expenses that will be incurred in servicing insurance obligations. The best estimate has to be evaluated for each line of business. The market-wide estimates of the net of reinsurance standard deviation for reserve risk are given for each line of business. For each one, standard deviation coefficient corresponds to the standard deviation of $\frac{\hat{R}^{I+1}_i + X_{i,I-i+1}}{\hat{R}^I_i}$. The calibration of these coefficients is to date still the purpose of debates and analysis.

**Standard formula with « *undertaking-specific* » parameters**

Undertaking-specific parameters are an important element of the standard formula: they contribute to more risk-sensitive capital requirements and facilitate the risk management of undertakings. Subject to the approval of the supervisory authorities, insurance and reinsurance undertakings may, within the design of the standard formula, replace the standard deviation for non-life premium risk and the standard deviation for non-life reserve risk, for each segment.

In any application for approval of the use of undertaking-specific parameters to replace a subset of parameters of the standard formula, insurance and reinsurance undertakings shall in particular demonstrate that standard formula parameters do not reflect the risk profile of the company.

Such parameters shall be calibrated on the basis of the internal data of the undertaking concerned, or of data which is directly relevant for the operations of that undertaking using standardized methods. When granting supervisory approval, supervisory authorities shall verify the completeness, accuracy and appropriateness of the data used.

Insurance and reinsurance undertakings shall calculate the standard deviation of the undertaking by using, for each parameter, a standardized method.

Regarding the reserve risk in the QIS5 exercise, 3 standardized methods have been defined for the calibration of undertaking-specific parameters:
- The first one corresponds to a retrospective approach, based on the volatility of historical economic boni or mali,
- The two other methods are based on the model of Wüthrich *et al.* (2008).

In the context of the QIS 5 calculation, the data used should meet a set of binding requirements. In particular the estimation should be made on complete claims triangles for payments.

Many undertakings have also expressed the wish of a wider panel of standardized methods which includes simulation methods such as those proposed in this study.

*Internal model*

For one accident year $i$, the CDR has been defined as follows:
$$\widehat{CDR}_i(I+1) = \hat{C}^I_{i,I_{ult}} - \hat{C}^{I+1}_{i,I_{ult}} = \hat{R}^I_i - \left(\hat{R}^{I+1}_i + X_{i,I-i+1}\right).$$

This CDR corresponds to the difference between two successive predictions of the total ultimate claim.

The capital requirement for the reserve risk, for accident year $i$ is given by:
$$SCR^i_{réserve} = -VaR_{0,5\%}\left(\widehat{CDR}_i(I+1)\right).$$

For all accident years:
$$SCR_{réserve} = -VaR_{0,5\%}\left(\sum_{i=1}^{I}\widehat{CDR}_i(I+1)\right) = -VaR_{0,5\%}\left(\sum_{i=1}^{I}\widehat{CDR}(I+1)\right).$$

When using an internal model, the goal is to evaluate the 0.5%-percentile of the $\widehat{CDR}(I+1)$ distribution. This allows to calculate the capital requirement for the reserve risk in a stand-alone approach (before aggregation with other risks).

# 3. State of the art of one-year stochastic reserving methods

This section deals with the analytical results of Wüthrich *et al.* (2008) and existing simulation methods providing an empirical distribution of the next-year CDR. The results shown in this paper are applied to a loss development triangle with incremental payments $(X_{i,j}, 0 \leq i + j \leq I)$ or cumulative payments $(C_{i,j}, 0 \leq i + j \leq I)$, illustrated by Figure 2. So we suppose in this paper that the number of accident years is equal to the number of development years, and the results are presented in this context. Finally, we will not deal with inflation. However, the adaptation of the methods thereafter presented with the aim of its integration raises no theoretical problem.

|  | Development year *j* | | | | |
|---|---|---|---|---|---|
| **Accident year *i*** | **0** | **1** | ***j*** | ***I* − 1** | ***I*** |
| **0** | $C_{0,0}$ | $C_{0,1}$ | … | $C_{0,I-1}$ | $C_{0,I}$ |
| **1** | $C_{1,0}$ | $C_{1,1}$ | … | $C_{1,I-1}$ | |
| ***i*** | … | … | … | | |
| ***I* − 1** | $C_{I-1,0}$ | $C_{I-1,1}$ | | | |
| ***I*** | $C_{I,0}$ | | | | |

**Figure 2:** Loss development triangle of cumulative payments.

## 3.1. Analytical results on the volatility of the CDR

The aim of the paper of Wüthrich *et al.* (2008) is to quantify the uncertainty linked to the re-evaluation of the best estimate between time $I$ and $I + 1$. We refer to Wüthrich *et al.* (2008) for the proof of the presented results, which are used thereafter.

### 3.1.1. Model assumptions

The time series model of Wüthrich *et al.* (2008) has been proposed by Buchwalder *et al.* (2006) and is based on the assumptions below:
- The cumulative payments $C_{ij}$ in different accident years $i \in \{0, ..., I\}$ are independent.
- There exist constants $f_l > 0$ and $\sigma_l > 0$, $l \in \{0, ..., I - 1\}$, such that for all $j \in \{1, ..., I\}$ and for all $i \in \{0, ..., I\}$,

$$C_{i,j} = f_{j-1} C_{i,j-1} + \sigma_{j-1} \sqrt{C_{i,j-1}}\, \epsilon_{i,j}.$$

- Given $B_0 = \{C_{0,0}, \ldots, C_{I,0}\}$ where for all $i \in \{0, \ldots, I\}, C_{i,0} > 0$, the random variables $\epsilon_{i,j}$ are independent with:

  $\forall i \in \{0, \ldots, I\}, \forall j \in \{1, \ldots, I\}, \mathbb{E}[\epsilon_{i,j}|B_0] = 0, \mathbb{E}[\epsilon_{i,j}^2|B_0] = 1$, and $\mathbb{P}[C_{i,j} > 0|B_0] = 1$.

The equations which follow are studied under measure $\mathbb{P}[.|B_0]$.

**Remark:** These assumptions defining a time series for each accident year $i$ are stronger than those underlying the conditional and non-parametric model of Mack (1993) characterizing only the first two moments of the cumulative payments $C_{i,j}$.

It is pointed out that the development factors estimated by the Chain Ladder method at time $I$ are given by

$$\forall j \in \{0, \ldots, I-1\}, \ \hat{f}_j^I = \frac{\sum_{i=0}^{I-j-1} C_{i,j+1}}{S_j^I},$$

with

$$S_j^I = \sum_{i=0}^{I-j-1} C_{i,j}.$$

At time $I + 1$, the Chain Ladder development factors take into account new information, i.e. observed cumulative payments in the sub-diagonal to come. These Chain Ladder factors are thus written

$$\forall j \in \{0, \ldots, I-1\}, \ \hat{f}_j^{I+1} = \frac{\sum_{i=0}^{I-j} C_{i,j+1}}{S_j^{I+1}},$$

with

$$S_j^{I+1} = \sum_{i=0}^{I-j} C_{i,j}.$$

In this context, the unbiased estimator of $(\sigma_j)^2$ proposed by Mack (1993) and in particular used by Wüthrich *et al.* (2008) is, for all $j \in \{1, \ldots, I-1\}$:

$$(\hat{\sigma}_{j-1}^I)^2 = \frac{1}{I-j} \sum_{i=0}^{I-j} C_{i,j-1} \left( \frac{C_{i,j}}{C_{i,j-1}} - \hat{f}_{j-1}^I \right)^2.$$

Moreover, the estimate of the last variance parameter can be done for example according to the approximation suggested by Mack (1993):

$$(\hat{\sigma}_{I-1}^I)^2 = \min \left( \frac{(\hat{\sigma}_{I-2}^I)^4}{(\hat{\sigma}_{I-3}^I)^2}, \min((\hat{\sigma}_{I-3}^I)^2, (\hat{\sigma}_{I-2}^I)^2) \right).$$

### 3.1.2. Claims Development Result

Within the framework of Solvency II, it is necessary to be able to measure the uncertainty related to the re-estimation of the best estimate between time $I$ and $I + 1$. Thus, the variable of interest is not any more the payments until the ultimate but the Claims Development Result defined in this part.

The re-estimation of the best estimate is based on two sets of information. Let $D_I$ denote the available information at time $I$ (i.e. the upper triangle). After one year, information $D_I$ is enlarged with the observation of the cumulative payments of the sub-diagonal to come. Thus, information known at time $I + 1$, denoted $D_{I+1}$, contains the original triangle and the sub-diagonal observed: we have thus $D_I \subset D_{I+1}$.

### 3.1.2.1. True CDR

Formally, the true Claims Development Result in accounting year $(I, I + 1]$ for accident year $i$ is defined in the following way:

$$CDR_i(I + 1) = E[R_i^I|D_I] - (X_{i,I-i+1} + E[R_i^{I+1}|D_{I+1}]),$$

where

- $R_i^I = C_{i,I} - C_{i,I-i}$ is the expected outstanding liabilities conditional on $D_I$ for accident year $i$,
- $R_i^{I+1} = C_{i,I} - C_{i,I-i+1}$ is the expected outstanding liabilities conditional on $D_{I+1}$ for accident year $i$,
- $X_{i,I-i+1} = C_{i,I-i+1} - C_{i,I-i}$ denotes the incremental payments between time $I$ and time $I + 1$ for accident year $i$.

The true CDR for aggregated accident years is given by

$$CDR(I + 1) = \sum_{i=1}^{I} CDR_i(I + 1).$$

We can decompose the true CDR of year $(I, I + 1]$ as the difference between two successive estimations of the expected ultimate claims, i.e.

$$CDR_i(I + 1) = E[C_{i,I}|D_I] - E[C_{i,I}|D_{I+1}].$$

### 3.1.2.2. Observable CDR

The true CDR defined previously is not observable because the "true" Chain Ladder factors are unknown. The CDR which is based on an estimation of the expected ultimate claims by the Chain Ladder method is called "observable CDR". It represents the position observed in the income statement at time $I + 1$ and is defined in the following way:

$$\widehat{CDR}_i(I + 1) = \hat{R}_i^{D_I} - (X_{i,I-i+1} + \hat{R}_i^{D_{I+1}}),$$

where $\hat{R}_i^{D_I}$ (resp. $\hat{R}_i^{D_{I+1}}$) is the Chain Ladder estimator of $E[R_i^I|D_I]$ (resp. $E[R_i^I|D_{I+1}]$), the ultimate claims expected value at time $I$ (resp. $I + 1$). These estimators are unbiased conditionally to $C_{i,I-i}$.

In the same way, the aggregate observable CDR for all accident years is defined by

$$\widehat{CDR}(I + 1) = \sum_{i=1}^{I} \widehat{CDR}_i(I + 1).$$

We will present, in the continuation, the error of prediction of the true CDR by the observable CDR at time $I$ and we will be more particularly interested in the error of prediction of the observable CDR by value 0 at time $I$.

### 3.1.2.3. *Errors of prediction*

The analytical results of Wüthrich *et al.* (2008) relate to two errors of prediction :
- The prediction of the true CDR by the observable CDR at time $I$,
- The prediction of the observable CDR by 0 at time $I$.

The conditional Mean Square Error of prediction (MSE) of the true CDR by the observable CDR, given $D_I$, is defined by

$$MSE_{D_I}\left(\sum_{i=1}^{I} \widehat{CDR}_i(I+1)\right) = \mathbb{E}\left[\left(\sum_{i=1}^{I} \widehat{CDR}_i(I+1) - \sum_{i=1}^{I} CDR_i(I+1)\right)^2 \bigg| D_I\right].$$

This error represents the conditional average quadratic difference between the true CDR and the observable CDR, given information $D_I$ contained in the upper triangle.

As for the conditional mean square error of prediction of the observable CDR by 0, given $D_I$, it is defined by

$$\mathbb{E}\left[\left(\sum_{i=1}^{I} \widehat{CDR}_i(I+1) - 0\right)^2 \bigg| D_I\right].$$

This error, homogeneous to a moment of order 2, is linked to a prospective vision: which error does one make by predicting the observable CDR, which will appear in the income statement at the end of the accounting year $(I, I+1]$, by value 0 at time $I$ ? The quantification of this error is the object of the following developments

### 3.1.3. Error of prediction of the true CDR by the observable CDR

For accident year $i$, the conditional mean square error of prediction of the true CDR by the observable CDR is

$$MSE_{D_I}\left(\widehat{CDR}_i(I+1)\right) = \Phi_{i,I}^I + \left(\mathbb{E}[\widehat{CDR}_i(I+1) | D_I]\right)^2, \quad (3.5)$$

with

$$\Phi_{i,I}^I = Var\left(CDR_i(I+1) - \widehat{CDR}_i(I+1) | D_I\right)$$
$$= Var(CDR_i(I+1)|D_I) + Var(\widehat{CDR}_i(I+1)|D_I)$$
$$- 2\, Cov\left(\widehat{CDR}_i(I+1), CDR_i(I+1) | D_I\right). \quad (3.6)$$

The first term is the process error of the true CDR and the second can be interpreted as the process error of the observable CDR (the CDR observed at time $I+1$ is a random variable at time $I$). The last term (covariance) takes into account the correlation between the true CDR and the observable CDR.

**Estimation of the first moment for accident year $i$**

The expression $\left(E[\widehat{CDR}_i(I+1) | D_I]\right)^2$ is related to the bias of the observable CDR as an estimator of the true CDR. In general, the estimation error quantifies the distance between an unknown parameter and the estimator proposed to approach this parameter. Here, the existence of this term comes from the estimation error made by approaching the unknown development factors by the Chain Ladder factors. Wüthrich *et al.* (2008) proposes the estimator $\left(\hat{C}_{i,j}\right)^2 \widehat{\Delta}_{i,I}^I$ of the (mean square) estimation error, with

$$\widehat{\Delta}_{i,I}^{I} = \frac{(\hat{\sigma}_{I-i}^{I})^{2}}{(\hat{f}_{I-i}^{I})^{2} S_{I-i}^{I}} + \sum_{j=I-i+1}^{I-1} \left(\frac{C_{I-j,j}}{S_{j}^{I+1}}\right)^{2} \frac{(\hat{\sigma}_{j}^{I})^{2}}{(\hat{f}_{j}^{I})^{2} S_{j}^{I}}. \quad (3.10)$$

**Estimation of the second moment for accident year $i$**

The (mean square) process error for accident year $i$ corresponding to the equation (3.6) is estimated by Wüthrich *et al.* (2008) in the following way:

$$\widehat{\Phi}_{i,I}^{I} = (\hat{C}_{i,I})^{2} \left[ 1 + \frac{(\hat{\sigma}_{I-i}^{I})^{2}}{(\hat{f}_{I-i}^{I})^{2} C_{i,I-i}} \right] \left[ \prod_{l=I-i+1}^{I-1} \left( 1 + \frac{(\hat{\sigma}_{l}^{I})^{2}}{(\hat{f}_{l}^{I})^{2} (S_{l}^{I+1})^{2}} C_{I-l,l} \right) - 1 \right]. \quad (3.9)$$

Finally, Wüthrich *et al.* (2008) proposes the following estimator of the (mean square) error of prediction for each accident year $i$:

$$\widehat{MSE}_{D_I}\left(\widehat{CDR}_i(I+1)\right) = \widehat{\Phi}_{i,I} + (\hat{C}_{i,I})^2 \, \widehat{\Delta}_{i,I}^I.$$

The mean square error of prediction for aggregated accident years is estimated by

$$\widehat{MSE}_{D_I}\left(\sum_{i=1}^{I} \widehat{CDR}_i(I+1)\right) = \sum_{i=1}^{I} \widehat{MSE}_{D_I}\left(\widehat{CDR}_i(I+1)\right) + 2 \sum_{i>k>0} \left(\widehat{\Psi}_{i,k}^{I} + \hat{C}_{i,I}^{I} \hat{C}_{k,I}^{I} \widehat{\Lambda}_{k,I}^{I}\right),$$

with, for $i > k > 1$,

$$\widehat{\Psi}_{i,k}^{I} = \frac{\hat{C}_{i,I}^{I}}{\hat{C}_{k,I}^{I}} \left(1 + \frac{(\hat{\sigma}_{I-k}^{I})^{2}}{(\hat{f}_{I-k}^{I})^{2} S_{I-k}^{I+1}}\right) \left(1 + \frac{(\hat{\sigma}_{I-k}^{I})^{2}}{(\hat{f}_{I-k}^{I})^{2} C_{k,I-k}}\right)^{-1} \widehat{\Phi}_{k,I}^{I},$$

$$\widehat{\Lambda}_{k,I}^{I} = \frac{C_{k,I-k}}{S_{I-k}^{I+1}} \frac{(\hat{\sigma}_{I-k}^{I})^{2}}{(\hat{f}_{I-k}^{I})^{2} S_{I-k}^{I}} + \sum_{j=I-k+1}^{I-1} \left(\frac{C_{I-j,j}}{S_{j}^{I+1}}\right)^{2} \frac{(\hat{\sigma}_{j}^{I})^{2}}{(\hat{f}_{j}^{I})^{2} S_{j}^{I}}, \quad (3.13)$$

and $\widehat{\Psi}_{i,1}^{I} = 0$ for $i > 1$.

### 3.1.4. Error of prediction of the observable CDR by 0

The estimator of the error of prediction of the observable CDR by 0 proposed by Wüthrich *et al.* (2008) partly refers to previously presented expressions. This error of prediction is defined by

$$\mathbb{E}\left[\left(\sum_{i=1}^{I} \widehat{CDR}_i(I+1) - 0\right)^2 \bigg| D_I\right]$$

$$= \underbrace{\widehat{\mathbb{E}}_{D_I}\left[\left(\mathbb{E}\left[\sum_{i=1}^{I} \widehat{CDR}_i(I+1) - 0 \bigg| D_I\right]\right)^2\right]}_{(u-bias)^2} + \widehat{Var}\left(\sum_{i=1}^{I} \widehat{CDR}_i(I+1) \bigg| D_I\right). (3.15)$$

The error of prediction for aggregated accident years is decomposed as follows:
- On the one hand, the term "$(u - bias)^2$" of Wüthrich *et al.* (2008) is the expected quadratic existing bias between the observable CDR and its estimator (the value 0). This term can be interpreted as the overall estimation error linked to this prediction at time $I$.
- On the other hand, the variance of the observable CDR is the process error related to this prediction. This inevitable error results from the randomness of the variable to be predicted (here the observable CDR).

The aggregated estimation error breaks up into estimation errors for each accident year and terms linked to the correlation between accident years in the following way:

$$(u - bias)^2 = \sum_{i=1}^{I} (\hat{C}_{i,I}^I)^2 \widehat{\Delta}_{i,I}^I + 2 \sum_{I \geq i > k \geq 1} \hat{C}_{i,I}^I \hat{C}_{k,I}^I \widehat{\Lambda}_{k,I}^I. \quad (3.14)$$

In the same way, the estimated aggregated process error is written

$$\widehat{Var}\left(\sum_{i=1}^{I} \widehat{CDR}_i(I+1) \bigg| D_I\right) = \sum_{i=1}^{I} \hat{\Gamma}_{i,I}^I + 2 \sum_{I \geq i > k \geq 1} \hat{Y}_{i,k}^I. \quad (3.16)$$

The estimator of the process error proposed by Wüthrich *et al.* (2008) for accident year $i \geq 1$ is given by

$$\hat{\Gamma}_{i,I}^I = \widehat{Var}(\widehat{CDR}_i(I+1)|D_I)$$
$$= (\hat{C}_{i,I}^I)^2 \left\{ \left(1 + \frac{(\hat{\sigma}_{I-i}^I)^2}{(\hat{f}_{I-i}^I)^2 C_{i,I-i}}\right) \prod_{l=I+1-i}^{I-1} \left(1 + \frac{(\hat{\sigma}_l^I)^2 C_{I-l,l}}{(\hat{f}_l^I)^2 (S_l^{I+1})^2}\right) - 1 \right\}, \quad (3.17)$$

and for $i > k > 0$, the following covariance terms are given:

$$\hat{Y}_{i,k}^I = \widehat{Cov}(\widehat{CDR}_i(I+1), \widehat{CDR}_k(I+1)|D_I)$$
$$= \hat{C}_{i,I}^I \hat{C}_{k,I}^I \left\{ \left(1 + \frac{(\hat{\sigma}_{I-k}^I)^2}{(\hat{f}_{I-k}^I)^2 S_{I-k}^{I+1}}\right) \prod_{l=I+1-k}^{I-1} \left(1 + \frac{(\hat{\sigma}_l^I)^2 C_{I-l,l}}{(\hat{f}_l^I)^2 (S_l^{I+1})^2}\right) - 1 \right\}. \quad (3.18)$$

These analytical results are the subject of section 4.

### 3.2. One-year simulation methods

#### 3.2.1. Adaptation of ultimate simulation methods to one-year horizon

In this section we synthesize the main steps of simulation methods measuring the one-year reserve risk, having for references Ohlsson *et al.* (2008) and Diers (2008). We also mention the main methods used in practice for one-year reserve risk simulation raised in the study of AISAM-ACME (2007).

The three steps below present the obtaining of a distribution of one-year future payments and best estimate starting from a loss development triangle:

1) Calculation of best estimate at time $I$. This best estimate is regarded as determinist since calculated on realized data, thus known at time $I$, contained in the upper triangle.
2) Simulation of the one-year payments between time $I$ and time $I + 1$: they are the incremental payments in the sub-diagonal of the loss development triangle.
3) On the basis of step 2, calculation of the best estimate at time $I + 1$.

These steps are illustrated by Figure 3 below, extracted from Diers (2008):

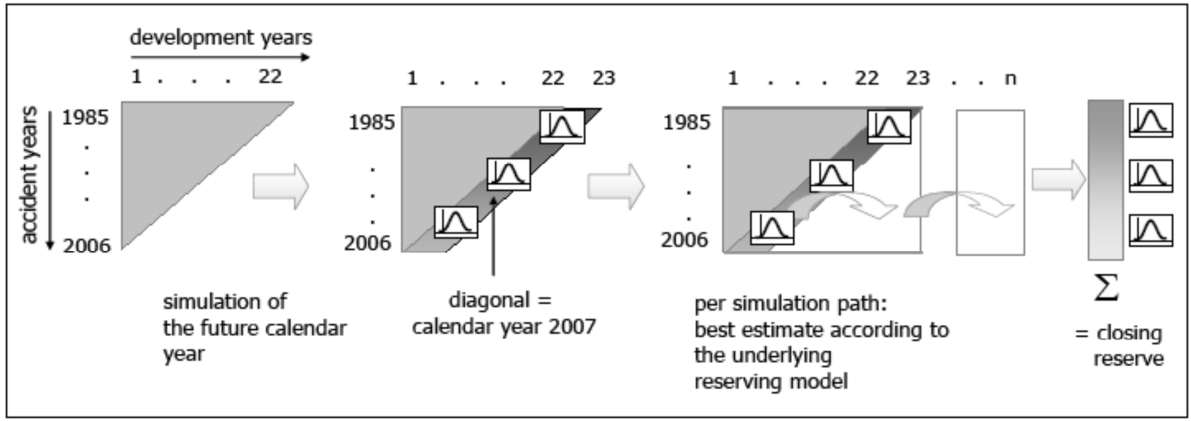

**Figure 3:** One-year reserve risk simulation.

In practice, these steps are used as a basis in order to adapt existing ultimate simulation methods to one-year horizon. One can in particular raise in AISAM-ACME (2007) the use of bootstrap methods using the time series model of Wüthrich *et al.* (2008) as well as the adaptation of Bayesian simulation methods to one-year horizon. We present in the following sections these two methods, and two possible adaptations of GLM bootstrap methods to measure one-year uncertainty for reserves.

### 3.2.2. Bayesian methods

The adaptation of the Bayesian simulation methodology to one-year horizon raised by AISAM-ACME (2007) is in particular based on the model suggested by Scollnik (2004). An adaptation to one-year horizon of this method is also mentioned by Lacoume (2008). We propose in this part a synthesis of the use of the Bayesian simulation method and its adaptation to the one-year horizon.

#### 3.2.2.1. *Model assumptions*

We present the Bayesian framework of the Chain Ladder model, developed by Scollnik (2004), which models the individual development factors. Based on the observation of the similar values of $(f_{i,j})_{0 \leq i \leq I-1}$ for the same development year *j*, the following model is suggested:
$f_{i,j} \sim N(\theta_j, \tau)$ where $\theta_j \sim N(\mu_\theta, \tau_\theta)$.

#### 3.2.2.2. *A priori distributions*

The characteristic of the Bayesian methodology is the specification of an *a priori* distribution for each model parameter, seen like a random variable (here $\tau$, $\mu_\theta$ et $\tau_\theta$). This *a priori* distribution can be informative (reduced variance) or non-informative (high variance). The non-informative *a priori* distributions are generally applied to several parameters, possibly having different characteristics, in a generic way. The choice between these two kinds of information does not return within the framework of this paper, and we refer to Scollnik (2004) for more details. The *a priori* distributions suggested by Scollnik (2004) are

$$\tau \sim \Gamma(a, b),$$
$$\mu_\theta \sim N(a', b'),$$
$$\tau_\theta \sim \Gamma(a'', b''),$$

where $a, b, a', b', a''$ and $b''$ are fixed parameters.

### 3.2.2.3. Calculation of the a posteriori distribution of the parameters

First, let $x = (\tau, \mu_\theta, \tau_\theta)$ denote the parameters of interest and $\hat{f}_{i,j}$ the observed individual development factors. The aim of this step is to provide a distribution having for density $\pi(x|\hat{f}_{i,j})$, the conditional probability of the parameters of interest given the observed data: it is the *a posteriori* distribution of the parameters. According to Bayes' theorem, this *a posteriori* density is written

$$\pi(x|\hat{f}_{i,j}) = \frac{\pi(\hat{f}_{i,j}|x)\pi(x)}{\int \pi(\hat{f}_{i,j}|x)\pi(x)dx},$$

where $\pi(\hat{f}_{i,j}|x)$ is the conditional density of the data, given the parameters of interest and $\pi(x)$ is the density of the parameters of interest (specification of *a priori* distributions).

### 3.2.2.4. Markov Chain Monte Carlo (MCMC) algorithm

To obtain a conditional empirical distribution of the parameters of interest, given the observed data, one uses MCMC technique. The aim of MCMC algorithm is to build a Markov chain whose stationary law is $\pi(x|\hat{f}_{i,j})$. Thus, it makes it possible to generate samples from the *a posteriori* law of the parameters of interest and we denote here $(x_t)_{1 \leq t \leq N}$ the distribution of parameters generated thanks to this procedure. We refer to Scollnik (2001) for a more detailed presentation of MCMC algorithm.

### 3.2.2.5. A posteriori distribution of the variable of interest

The final goal is to obtain an empirical distribution of future individual development factors $f_{i,j}$. The conditional distribution of the future development factors, given the observed data, is written

$$\pi(f_{i,j}|\hat{f}_{i,j}) = \int \pi(f_{i,j}|x)\pi(x|\hat{f}_{i,j})dx.$$

The ergodic property of the Markov chain built by MCMC algorithm makes it possible to write

$$\pi(f_{i,j}|\hat{f}_{i,j}) \approx \frac{1}{N} \sum_{t=1}^{N} \pi(f_{i,j}|x_t).$$

Thus, to obtain samples of $f_{i,j}$ conditionally on the observed data, one simulates $f_{i,j}$ conditionally on the generated parameters $x_t$ for all $t \in \{1, \ldots, N\}$, thanks to the model expression.

### 3.2.2.6. Adaptation of this method to one-year horizon

The steps carrying out the adaptation to one-year horizon of Bayesian approaches are mentioned by Lacoume (2008) and AISAM-ACME (2007). We present here a synthesis of this adaptation in three steps:

1) Calculation of $BE_I$ using the original Chain Ladder factors.

**Iteration No $k$**

2) Simulation of the sub-diagonal to come using the Bayesian model: the method described above generates a realization of the next year development factors $\left(f_{i,I-i}^k\right)_{i\in\{1,\ldots,I\}}$, and then a realization of the payments of the sub-diagonal $P_{I+1}^k$ by

$$P_{I+1}^k = \sum_{i=1}^{I} f_{i,I-i}^k C_{i,I-i} - C_{i,I-i}.$$

3) Re-estimation of the Chain Ladder factors on the trapezoid and calculation of the best estimate at time $I+1$, $BE_{I+1}^k$.

**End of iteration No $k$**

The variance related to the revision of the best estimate between time $I$ and time $I+1$ is given by

$$Var\left(\left(P_{I+1}^k + BE_{I+1}^k\right)_{1\leq k\leq K}\right),$$

where $K$ is the sample size.

Thus, Bayesian Chain Ladder model is a simulation tool providing a one-year empirical distribution. Nevertheless, the use of this method is tricky because MCMC algorithm can diverge in certain cases. We refer to Scollnik (2004) and Lacoume (2008) for the implementation of the method using WinBUGS software and for practical problems.

### 3.2.3. One-year GLM bootstrap

We present in this section two existing adaptations of GLM (Generalized Linear Models) bootstrap methodologies, generating samples of the CDR in a one-year view. Historically, the generalized linear models have been proposed by J. Nelder and R. Wedderburn in 1972 and in the field of non-life reserving, Renshaw *et al.* (1994, 1998) have studied log-Poisson model reproducing results of the Chain Ladder method while generating a full empirical distribution of future payments.

This bootstrap method is applied to a loss development triangle with incremental payments $\left(X_{i,j}\right)_{0\leq i+j\leq I}$. We present here the particular case of a GLM model with Over-Dispersed Poisson (ODP) distribution. Within this framework, the incremental payments $X_{i,j}$ are modelled with an ODP distribution with mean $\mu_{i,j}$ and variance $\Phi\mu_{i,j}^2$, where

$$\mathbb{E}[X_{i,j}] = \mu_{i,j} = \exp(\mu + \alpha_i + \beta_j).$$

#### 3.2.3.1. First approach

In the context of the adaptation of these methods to measure one-year reserve risk, this first approach has been proposed by Boisseau (2010) and Lacoume (2008). This method, which is applied to the residuals of the incremental payments of the loss development triangle, is described below.

**Step 1**

**1.a.** Estimation of the model parameters $\hat{\mu}, (\hat{\alpha}_i)_{0\leq i\leq I}$, and $(\hat{\beta}_j)_{0\leq j\leq I}$ on the upper triangle and calculation of the expected values $(\hat{\mu}_{i,j})_{0\leq i+j\leq I}$.

**1.b.** Calculation of best estimate at time $I$ by

$$BE_I = \sum_{i+j\geq I+1} \exp(\hat{\mu} + \hat{\alpha}_i + \hat{\beta}_j).$$

**1.c.** Calculation of the scale parameter $\hat{\phi} = \frac{\Sigma(r_{i,j}^P)^2}{n-p}$ where

- $n$ is the number of incremental payments in the upper triangle,
- $p$ is the number of parameters,
- $r_{i,j}^P = \frac{y_{i,j} - \hat{\mu}_{i,j}}{\sqrt{V(\hat{\mu}_{i,j})}}$ is the Pearson residual of $X_{i,j}$.

**1.d.** Calculation of the adjusted residuals by $r_{i,j}^A = \sqrt{\frac{n}{n-p}} r_{i,j}^P$.

**Iteration No $b$**

**Step 2**

Resampling with replacement of the residuals $r_{i,j}^A$ and construction of a pseudo-triangle of values $(X_{i,j}^b)_{0\leq i+j\leq I}$.

**Step 3**

Re-estimation of the model parameters and calculation of the following new parameters:
$\hat{\mu}^{b,1}, (\hat{\alpha}_i^{b,1})_{0\leq i\leq I}$ and $(\hat{\beta}_j^{b,1})_{0\leq j\leq I}$.

**Step 4**

Calculation of expected values $(\hat{\mu}_{i,j}^{b,1})_{i+j=I+1}$ in the sub-diagonal to come.

**Step 5**

Taking into account of process error. We obtain an incremental payment $X_{i,j}^b$ by simulation from a law with mean $\hat{\mu}_{i,j}^{b,1}$ and variance $\hat{\phi} (\hat{\mu}_{i,j}^{b,1})^2$. The one-year future payments are then

$$P_{I+1}^b = \sum_{i+j=I+1} X_{i,j}^b.$$

**Step 6**

Re-estimation of the model parameters $\hat{\mu}^{b,2}, (\hat{\alpha}_i^{b,2})_{0\leq i\leq I}$ and $(\hat{\beta}_j^{b,2})_{0\leq j\leq I}$ on the trapezoid $\{(X_{i,j}^b)_{0\leq i+j\leq I} \cup (X_{i,j}^b)_{i+j=I+1}\}$ in order to calculate new expected values $(\hat{\mu}_{i,j}^{b,2})_{i+j\geq I+2}$ in the following diagonals.

**Step 7**

Calculation of best estimate at time $I+1$ by

$$BE_{I+1}^b = \sum_{i=2}^{I} \sum_{j=I-i+2}^{I} \hat{\mu}_{i,j}^{b,2}.$$

We can notice here that only the expected incremental payments in the lower triangle, except the sub-diagonal, are required for the best estimate calculation at time $I+1$.

**End of iteration No $b$**

This method provides:
- An empirical distribution of one-year future payments $P_{I+1}^b$ in the sub-diagonal $(I, I+1]$,
- An empirical distribution of the best estimate at time $I+1$, $BE_{I+1}^b$.

We deduct then an empirical distribution of the CDR by
$$CDR^b = BE_I - P_{I+1}^b - BE_{I+1}^b.$$
The variance of this empirical distribution provides the prediction error, to compare with the prediction error (3.15) of Wüthrich *et al.* (2008).

### *3.2.3.2. Improvement of this method*

#### 3.2.3.2.1. Limits of the previous approach

The approach previously detailed raises two statistical issues. The statistical aspects mentioned below are raised and detailed by Boisseau (2010).

**Independence of the random variables**
Step 4 provides expected values in the sub-diagonal starting from the GLM parameters estimated on the upper triangle. Thus, at step 6 of re-estimation of the GLM parameters on the trapezoid by maximum likelihood, the random variables of the upper triangle and those of the sub-diagonal are not independent. The framework of maximum likelihood estimation, in which total probability breaks up into product of probabilities of the incremental payments, is thus not verified.

**Estimation error**
This method estimates the GLM parameters twice (steps 3 and 6): first to obtain incremental payments in the sub-diagonal and second to calculate the expected future payments (lower triangle). This approach tends to significantly increase the estimation error compared to the result of Wüthrich *et al.* (2008), and thereafter the total variance.

The second approach thereafter suggested makes it possible to overcome these limits.

#### 3.2.3.2.2. Steps of the improved bootstrap GLM procedure

This improvement has been proposed by Boisseau (2010). In each iteration of this new approach, the residuals of the original triangle are resampled on the trapezoid containing the upper triangle and the sub-diagonal. We present below the steps of this procedure.

**Step 1** is identical to that of the first approach.

**Iteration No** $b$

**Step 2**
Resampling with replacement of the residuals in the trapezoid and calculation of a pseudo-trapezoid of incremental payments $X_{i,j}^b$. This provides a *realization* of the payments of the sub-diagonal (taking into account of process error) from *resampled residuals* (taking into account of estimation error).

**Step 3**
Re-estimation of the model parameters on the pseudo-trapezoid. This provides new parameters $\hat{\mu}^b, (\hat{\alpha}_i^b), (\hat{\beta}_j^b)$ and new expected values $\hat{\mu}_{i,j}^b$ in the following diagonals starting from year $I + 1$.
We then calculate the best estimate at time $I + 1$ by :

$$\sum_{i=2}^{I} \sum_{j=I-i+2}^{I} \hat{\mu}_{i,j}^b .$$

**End of iteration No** $b$

The variance of the distribution provides the prediction error linked to the CDR calculation.

### 3.2.4. One-year recursive bootstrap method

The boostrap method proposed by De Felice *et al.* (2006) using the conditional resampling version of Mack model introduced by Buchwalder *et al.* (2006), is the basis of the study in section 4. This one-year simulation method is also mentioned by AISAM-ACME (2007) and Diers (2008).

This method provides a full empirical distribution of ultimate future payments (*Liability-at-Maturity approach*) and of one-year payments and best estimate (*Year-End-Expectation approach*), whose variance reproduces closed-form expressions proposed by De Felice *et al.* (2006). These expressions, with no taking into account of any discount effect, are equivalent to the estimators obtained by Wüthrich *et al.* (2008) related to the prediction error of the observable CDR by 0.
This one-year recursive bootstrap method resamples the residuals of the individual development factors, and can be applied to the residuals of the cumulative payments in the same way.

In the following, we detail the steps of this method and propose the inclusion of a tail factor. We also propose the proofs of equivalence of the variance of the simulated empirical distribution and analytical results of Wüthrich *et al.* (2008), and propose closed-form expressions including the stochastic modeling of the tail factor.

# 4. One-year recursive bootstrap method and inclusion of a tail factor

## 4.1. Introduction

The bootstrap method presented in this section provides an empirical distribution of the CDR whose variance replicates the prediction error of Wüthrich *et al.* (2008) mentioned in 3.1.4. In particular, this method makes it possible to include a tail factor simulated in each bootstrap iteration. Two alternatives of this method, providing the estimation error on the one hand and the process error on the other hand, are proposed in this section. Proofs of equivalence with the existing estimators are developed, and we also propose closed-form expressions including a tail factor.

This one-year bootstrap methodology including a tail factor is motivated by the need for replicating the analytical results of Wüthrich *et al.* (2008) used by CEIOPS for the calibration of reserve risk and proposed to date as a possible method for the "undertaking-specific" calibration. Indeed, the bootstrap methodology proposed in this section replicates existing closed-form expressions, while overcoming the limits of such an approach. In fact, deriving a full empirical distribution from the first two moments measurement proposed by Wüthrich *et al.* (2008) or splitting the distribution of the CDR into one-year payments and best estimate calculation in one year is not possible without additional assumption. These limits make it difficult to integrate the reserve risk in an internal model for example. One can also notice that analytical results of Wüthrich *et al.* (2008) do not include a tail development factor.

We present below the main advantages and the possible extensions of the bootstrap methodology proposed in this section.

- This method provides a full empirical distribution of the CDR and is thus not restricted to the calculation of the first two moments. It also measures reserve risk without assumption on the distribution of the CDR and provides a split between one-year payments and best estimate calculation in one year. Therefore, the inclusion of this method in an internal model taking into account other risks is direct.

- This one-year extension includes a stochastic modeling of the tail factor. Its use is therefore not restricted to loss triangles that are completely developed, which can be useful for lines of business with long development or having a lack of historical data.

- The calculation of the empirical distribution of the CDR allows also to take into account the whole dependency structure between lines of business, this one modeled for example by means of copulas.

- This method can be extended to measure the variability of the CDR in $K$ years, which is in particular useful within the ORSA (Own Risk and Solvency Assessment) framework. Here also, one will be able to obtain payments of year $(I + K, I + K + 1]$ on one hand, and best estimate at time $I + K + 1$ on the other hand.

- Lastly, this method makes it possible to take into account "management rules", i.e. the internal standards of the insurance company in terms of reserving policy. That can result, for example, in the exclusion of atypical individual development factors or the use of the Bornhuetter-Ferguson method to calculate the ultimate claim for the most recent accident years.

The study of the tail factor and the calculation of its variance are developed in 4.2. The steps of the bootstrap procedure are detailed in 4.3. We also propose a proof of equivalence between the variance of the empirical distributions and the estimators of the process and estimation errors suggested by Wüthrich *et al.* (2008) (see 4.4) as well as the closed-form expressions of the process and estimation errors including a tail factor (see 4.5). Finally, numerical results are shown in 4.6.

## 4.2. Inclusion of a tail factor

### 4.2.1. Extrapolation of the development factors

If the development of the triangle is not complete after $I$ development years, one can use a tail factor in order to estimate the ultimate payments at time $I_{ult} > I$. This tail development factor can be calculated by a linear extrapolation in the following way:

$$\forall j \in \{0, \ldots, I-1\}, \quad \ln(f_j - 1) = a.j + b,$$

with

$$f_{ult} = \prod_{j=I}^{I_{ult}-1} f_j.$$

### 4.2.2. Analytical estimate for the variance of the tail factor

In this section, we propose an analytical estimate $\sigma_{ult}^2$ for the variance of the tail factor related to the estimation error of the parameters $a$ and $b$ estimated by maximum likelihood technique. The linear extrapolation model at time $I$ can be written

$$Y = XZ,$$

with

$$Y = \begin{pmatrix} \ln(\hat{f}_0 - 1) \\ \vdots \\ \ln(\hat{f}_{I-1} - 1) \end{pmatrix},$$

$$X = \begin{pmatrix} 0 & 1 \\ \vdots & \vdots \\ I-1 & 1 \end{pmatrix},$$

and

$$Z = \begin{pmatrix} a \\ b \end{pmatrix}.$$

The maximum likelihood estimator $\hat{Z}$ is[1]

$$\hat{Z} = (\,^tXX)^{-1}\,^tXY.$$

The variance of the extrapolated tail factor is written

$$Var(\hat{f}_{ult}) = Var\left(\prod_{j=I}^{I_{ult}-1} \hat{f}_j\right) = Var\left(\prod_{j=I}^{I_{ult}-1}(1+\exp(\hat{a}.j+\hat{b}))\right) = Var\left(H_{I_{ult}-1}(\hat{Z})\right),$$

with

$$H_{I_{ult}-1}(\hat{Z}) = \prod_{j=I}^{I_{ult}-1}(1+\exp(\hat{a}.j+\hat{b})).$$

This allows to calculate the variance by means of the Delta method:

$$\widehat{Var}(\hat{f}_{ult}) \approx \,^t\nabla H_{I_{ult}-1}(\hat{Z})\,\widehat{\Sigma(\hat{Z})}\,\nabla H_{I_{ult}-1}(\hat{Z}),$$

where $\widehat{\Sigma(\hat{Z})}$ is an estimator of the variance-covariance matrix of $\hat{Z}$.

Thus, to diffuse the uncertainty related to the estimation of the parameters $a$ and $b$, one can simulate a tail development factor with mean

$$\hat{f}_{ult} = \prod_{j=I}^{I_{ult}-1} \hat{f}_j,$$

and variance $\sigma_{ult}^2 = \,^t\nabla H_{I_{ult}-1}(\hat{Z})\,\widehat{\Sigma(\hat{Z})}\,\nabla H_{I_{ult}-1}(\hat{Z})$ in each bootstrap iteration.

### 4.2.2.1. Calculation of the variance-covariance matrix

The maximum likelihood estimator $\hat{Z}$ being efficient, $\widehat{\Sigma(\hat{Z})}$ is the inverse of the estimated Fisher information matrix $\hat{I}$:

$$\widehat{\Sigma(\hat{Z})} = \hat{I}^{-1}.$$

The inverse of the estimated Fisher information matrix is written

$$\hat{I}^{-1} = \hat{\sigma}^2(\,^tXX)^{-1},$$

with

$$\hat{\sigma}^2 = \frac{1}{I}\sum_{j=0}^{I-1}\left(ln(\hat{f}_j-1)-\hat{a}j-\hat{b}\right)^2,$$

the biased estimator of the variance of the residuals.

### 4.2.2.2. Calculation of $\nabla H_{I_{ult}-1}$

The tail development factor extrapolated by parameters $a$ and $b$ is written

$$H_{I_{ult}-1}(a,b) = \prod_{j=I}^{I_{ult}-1}(1+\exp(a.j+b)).$$

In order to reduce the formulas, we propose here a recursive expression of $\nabla H_{I_{ult}-1}(a,b)$.

---

[1] In the framework of standard linear model in which residuals are supposed gaussian, the maximum likelihood estimator of $Z$ is the same as the least square estimator, whose expression is given here.

**Partial derivative with respect to $a$**

Let $j \geq 2$. We have
$$H_{I+j}(a,b) = \left(1 + e^{a(I+j)+b}\right) H_{I+j-1}(a,b),$$
then
$$\frac{\partial H_{I+j}}{\partial a}(a,b) = (I+j)e^{a(I+j)+b} H_{I+j-1}(a,b) + \left(1 + e^{a(I+j)+b}\right)\frac{\partial H_{I+j-1}}{\partial a}(a,b).$$
The original case is $\frac{\partial H_I}{\partial a}(a,b) = I e^{aI+b}$.

**Partial derivative with respect to $b$**

Let $j \geq 2$. We have
$$\frac{\partial H_{I+j}}{\partial b}(a,b) = e^{a(I+j)+b} H_{I+j-1}(a,b) + \left(1 + e^{a(I+j)+b}\right)\frac{\partial H_{I+j-1}}{\partial b}(a,b),$$

with $\frac{\partial H_I}{\partial b}(a,b) = e^{aI+b}$.

These two results give the recursive expression of the gradient of function $H_{I_{ult}-1}$:
$$\nabla H_{I_{ult}-1}(a,b) = \begin{pmatrix} \frac{\partial H_{I_{ult}-1}}{\partial a}(a,b) \\ \frac{\partial H_{I_{ult}-1}}{\partial b}(a,b) \end{pmatrix}.$$

Finally, we express once again the calculation of the variance of the tail factor by means of the Delta method:
$$\widehat{Var}(\hat{f}_{ult}) \approx {}^t\nabla H_{I_{ult}-1}(\hat{Z}) \, \widehat{\Sigma(\hat{Z})} \, \nabla H_{I_{ult}-1}(\hat{Z}). \quad (*)$$

### *4.2.3. Assumptions on the distribution of the tail factor*

$\hat{f}_{ult}$ is a function of the maximum likelihood estimator (MLE) $(\hat{a}, \hat{b})$. By invariance of this one by functional transformation, $\hat{f}_{ult}$ is the MLE of
$$f_{ult} = \prod_{j=I}^{I_{ult}-1} (1 + e^{aj+b}).$$
Within an asymptotic framework, $\hat{f}_{ult}$ is thus gaussian. We can then model the tail factor by a normal distribution with mean $\hat{f}_{ult}$ and variance $\sigma^2_{ult}$.

Nevertheless, it is also possible to adopt a more prudent approach by simulating a log-normal distribution as generally done in practice.

## 4.3. Description of the bootstrap procedure

We detail and illustrate the one-year bootstrap algorithm in 4.3.1, including the simulation of a tail factor. The remarks allowing to characterize the links between the simulation method and the analytical results are proposed in 4.3.2.

### 4.3.1. Steps of the bootstrap algorithm

This method carries out the resampling of the individual development factors residuals. The steps of this method are described below: step 1 is the original step carried out only once while steps 2 to 7 are a bootstrap iteration.

**Step 1**

**1.a.** Estimation of individual development factors $(f_{i,j})_{0 \leq i+j \leq I-1}$ and parameters $(\hat{f}_j)_{0 \leq j \leq I-1}$ and $(\hat{\sigma}_j)_{0 \leq j \leq I-1}$ on the original triangle of cumulative payments $(C_{i,j})_{0 \leq i+j \leq I}$.

**1.b.** Calculation of the expected tail development factor by extrapolation of the Chain Ladder development factors:

$$\hat{f}_{ult} = \prod_{j=I}^{I_{ult}-1} \hat{f}_j = \prod_{j=I}^{I_{ult}-1} (1 + e^{\hat{a}j+\hat{b}}).$$

**1.c.** Calculation of the best estimate $BE_I$ at time $I$ by

$$BE_I = \sum_{i=0}^{I} (\hat{C}_{i,I_{ult}} - C_{i,I-i}),$$

with

$$\hat{C}_{0,I_{ult}} = \hat{f}_{ult} C_{0,I},$$

and

$$\forall i \in \{1, \ldots, I\}, \hat{C}_{i,I_{ult}} = \hat{f}_{ult} \left( \prod_{j=I-i}^{I-1} \hat{f}_j \right) C_{i,I-i}.$$

**1.d.** Calculation of the residuals of the individual development factors by

$$\forall i,j \,/\, 0 \leq i + j \leq I - 1, \quad r_{i,j} = \frac{\sqrt{C_{i,j}}(f_{i,j} - \hat{f}_j)}{\hat{\sigma}_j}.$$

Residuals are then adjusted by

$$\forall i,j \,/\, 0 \leq i + j \leq I - 1, \quad r_{i,j}^A = \sqrt{\frac{I-j}{I-j-1}} \frac{\sqrt{C_{i,j}}(f_{i,j} - \hat{f}_j)}{\hat{\sigma}_j}.$$

Lastly, these residuals are centered. The adjustment by the factor $\sqrt{\frac{I-j}{I-j-1}}$ allows to correct the bias related to the calculation of the bootstrap variance, in order to make the analytical expression of the variance and the dispersion of the simulated distribution match.

**Iteration No $b$**

**Step 2**
Resampling with replacement of the residuals in the upper triangle and obtaining of an upper triangle of pseudo-development factors seen at time $I$:

$$\forall i,j \,/\, 0 \leq i + j \leq I - 1, \quad f_{i,j}^{b,I} = r_{i,j}^b \sqrt{\frac{\hat{\sigma}_j^2}{C_{i,j}}} + \hat{f}_j.$$

**Step 3**
Re-estimation of the Chain Ladder factors seen at time $I$ by

$$\forall j \in \{0, \ldots, I-1\}, \quad f_j^{b,I} = \frac{\sum_{i=0}^{I-j-1} C_{i,j} f_{i,j}^{b,I}}{\sum_{i=0}^{I-j-1} C_{i,j}}.$$

This is equivalent to calculate Chain Ladder factors by weighted average of the individual development factors, with weights equal to the cumulative payments of the original triangle.

**Step 4**

Simulation of the one-year payments in order to take into account process error. For all $i \in \{1, \ldots, I\}$, calculation of $C_{i,I+1-i}^b$ by simulating a normal distribution with mean $C_{i,I-i} f_{I-i}^{b,I}$ and variance $C_{i,I-i} (\hat{\sigma}_{I-i}^I)^2$. One deduces from it the future payments in next accounting year $(I, I+1]$ by

$$P_{I+1}^b = \sum_{i=1}^{I} (C_{i,I+1-i}^b - C_{i,I-i}).$$

**Step 5**

**5.a.** Calculation of new individual development factors $\left(f_{I-j,j}^{b,I+1}\right)_{0 \leq j \leq I-1}$ on the simulated sub-diagonal, and calculation of new Chain Ladder factors at the end of year $(I, I+1]$. These new Chain Ladder factors are estimated by the cumulative payments of the original triangle (information $D_I$) and the new individual development factors in the following way:

$$\forall j \in \{0, \ldots, I-1\}, \quad f_j^{b,I+1} = \frac{\sum_{i=0}^{I-j-1} C_{i,j} f_{i,j} + C_{I-j,j} f_{I-j,j}^{b,I+1}}{\sum_{i=0}^{I-j} C_{i,j}}.$$

**5.b.** Taking into account of the estimation error of the extrapolation parameters by simulating a tail development factor with mean $\hat{f}_{ult}$ and variance $\sigma_{ult}^2$.

**Step 6**

Calculation of the best estimate $BE_{I+1}^b$ seen at time $I+1$, starting from the simulated sub-diagonal $\left(C_{i,I-i+1}^b\right)_{1 \leq i \leq I}$, the pseudo-factors $\left(f_j^{b,I+1}\right)_{1 \leq j \leq I-1}$ and the tail factor $f_{ult}^b$ by

$$BE_{I+1} = \sum_{i=0}^{I} (C_{i,I_{ult}}^b - C_{i,I-i+1}^b),$$

with

$$C_{0,I_{ult}}^b = f_{ult}^b C_{0,I},$$
$$C_{1,I_{ult}}^b = f_{ult}^b C_{1,I}^b,$$

and

$$\forall i \in \{2, \ldots, I\}, C_{i,I_{ult}}^b = f_{ult}^b \left( \prod_{j=I-i+1}^{I-1} f_j^{b,I+1} \right) C_{i,I-i+1}^b.$$

**Step 7**

Calculation of the CDR of iteration No $b$:
$$CDR^b = BE_I - P_{I+1}^b - BE_{I+1}^b.$$

**End of iteration No $b$.**

### 4.3.2. Remarks

The previous algorithm allows to replicate the error of prediction of the observable CDR by 0 of Wüthrich *et al.* (2008) (see 4.4) and to obtain closed-form expressions including a tail factor (see 4.5). We emphasise here on the characteristics leading to such a result and propose the extensions allowing for the calculation of the two kinds of error.

- In **step 1**, we adopt the bias correction proposed by Mack (1993). The unbiased estimator of the variance parameter is written

$$(\hat{\sigma}_j^I)^2 = \frac{I-j}{I-j-1} \times \frac{1}{I-j} \sum_{i=0}^{I-j-1} C_{i,j} (f_{i,j} - \hat{f}_j^I)^2 = \frac{1}{I-j} \sum_{i=0}^{I-j-1} \left[ \sqrt{\frac{I-j}{I-j-1}} \sqrt{C_{i,j}} (f_{i,j} - \hat{f}_j) \right]^2.$$

The adjusted residuals are

$$\forall i,j \;/: 0 \leq i+j \leq I-1, \qquad r_{i,j}^A = \sqrt{\frac{I-j}{I-j-1}} \frac{\sqrt{C_{i,j}} (f_{i,j} - \hat{f}_j)}{\hat{\sigma}_j}.$$

These are the residuals included in the calculation of the scale parameter $(\hat{\sigma}_j^I)^2$. This adjustment causes the increase of the variance of the residuals and thus the variance of the empirical distribution, while leaving the mean of the residuals approximately unchanged (close to 0). This adjustment allows for the comparison with the analytical results and leads to very satisfactory relative differences (see section 4.6). Nevertheless, there are other possible bias corrections, proposed in particular by England *et al.* (1999,2002,2006) and Pinheiro *et al.* (2003).

- In **step 4**, the calculation of expected cumulative payments in the sub-diagonal, based on resampled pseudo-development factors, takes into account estimation error. Moreover, the simulation of these cumulative payments, given the variance parameters, includes process error: the calculated amounts are realizations of random variables. The taking into account of the two error components for the first sub-diagonal and the estimation error on the following diagonals is the same as in analytical results of Wüthrich *et al.* (2008) (eq. 3.17).

- At **step 5**, the re-estimation of Chain Ladder development factors at time $I + 1$ is done conditionally to the information in the trapezoid including the original triangle. This is written $D_I \subset D_{I+1}$, which is consistent with the CDR definition of Wüthrich *et al.* (2008) and compatible with the *« actuary-in-the-box »* point of view.

- The process and estimation errors including the tail development factor can be separately estimated by adapting this bootstrap procedure:

  - The process error is obtained by not carrying out neither the resampling of the residuals of the individual development factors, nor the tail factor simulation (we don't take into account the estimation error of the extrapolation parameters $a$ and $b$). This result is produced by deleting **steps 2 and 5.b**.

- The estimation error is obtained by not carrying out the simulation of the cumulative payments of the sub-diagonal, those being then seen as expected values: the process variance resulting from randomness of the cumulative payments is ignored. This modification corresponds to the calculation of $C_{i,I+1-i}^b = C_{i,I-i} f_{I-i}^{b,I}$ at **step 4**.

## 4.4. Proof of equivalence with the analytical results of Wüthrich *et al.* (2008)

In this section we prove that, with no tail factor, the variance of the distribution generated by the bootstrap procedure is equal to the prediction error of the observable CDR by 0 proposed by Wüthrich *et al.* (2008).

Let $\widehat{CDR}_i$ denote the CDR taking into account only estimation error and $CDR_i$ the CDR taking into account pure process error.

### 4.4.1. Estimation error

We neglect here process variance, therefore for all $i \in \{1, \dots, I\}$, $C_{i,I+1-i}^b = f_{I-i}^{b,I} C_{i,I-i}$.

#### 4.4.1.1. Estimation error for a single accident year

For $i \in \{1, \dots, I\}$, the variance of the CDR is written

$$Var(\widehat{CDR}_i) = Var\left( C_{i,I-i} \prod_{j=I-i}^{I-1} \hat{f}_j - f_{I-i}^{b,I} C_{i,I-i} \prod_{j=I+1-i}^{I-1} f_j^{b,I+1} \right).$$

**Remark:** Here and in all this study, an empty product is equal to 1, just as an empty sum is equal to 0.

The following results will be used thereafter, for $j \in \{0, \dots, I-1\}$:

- $f_j^{b,I} = \frac{\sum_{i=0}^{I-j-1} C_{i,j} f_{i,j}^{b,I}}{\sum_{i=0}^{I-j-1} C_{i,j}}$ avec $f_{i,j}^{b,I} = r_{i,j}^b \sqrt{\frac{\hat{\sigma}_j^2}{C_{i,j}}} + \hat{f}_j$,

- $\mathbb{E}[f_j^{b,I}] = \hat{f}_j$,

- $\mathbb{E}\left[\left(f_j^{b,I}\right)^2\right] = Var(f_j^{b,I}) + \hat{f}_j^2 = \frac{\hat{\sigma}_j^2}{S_j^I} + \hat{f}_j^2$,

- $f_j^{b,I+1} = \frac{\sum_{i=0}^{I-j-1} C_{i,j+1} + C_{I-j,j+1}^b}{\sum_{i=0}^{I-j} C_{i,j}} = \frac{\left(\sum_{i=0}^{I-j-1} C_{i,j}\right) \frac{\sum_{i=0}^{I-j-1} C_{i,j} f_{i,j}}{\sum_{i=0}^{I-j-1} C_{i,j}} + C_{I-j,j+1}^b}{\sum_{i=0}^{I-j} C_{i,j}} = \frac{S_j^I}{S_j^{I+1}} \hat{f}_j + f_j^{b,I} \frac{C_{I-j,j}}{S_j^{I+1}}$,

- $Var(f_j^{b,I+1}) = \left(\frac{C_{I-j,j}}{S_j^{I+1}}\right)^2 Var(f_j^{b,I}) = \left(\frac{C_{I-j,j}}{S_j^{I+1}}\right)^2 \frac{\hat{\sigma}_j^2}{S_j^I}$,

- $\mathbb{E}\left[\left(f_j^{b,I+1}\right)^2\right] = \left(\frac{C_{I-j,j}}{S_j^{I+1}}\right)^2 \frac{\hat{\sigma}_j^2}{S_j^I} + \hat{f}_j^2$.

Using the independence of the pseudo-development factors for different development years, we have

$$\mathbb{E}\left[C_{i,I-i}\prod_{j=I-i}^{I-1}\hat{f}_j - f_{I-i}^{b,I}C_{i,I-i}\prod_{j=I+1-i}^{I-1}f_j^{b,I+1}\right] = C_{i,I-i}\prod_{j=I-i}^{I-1}\hat{f}_j - \underbrace{\mathbb{E}[f_{I-i}^{b,I}]}_{\hat{f}_{I-i}}C_{i,I-i}\prod_{j=I+1-i}^{I-1}\underbrace{\mathbb{E}[f_j^{b,I+1}]}_{\hat{f}_j} = 0.$$

Thus, we have

$$Var(\widehat{CDR}_i) = \mathbb{E}\left[\left(C_{i,I-i}\prod_{j=I-i}^{I-1}\hat{f}_j - f_{I-i}^{b,I}C_{i,I-i}\prod_{j=I+1-i}^{I-1}f_j^{b,I+1}\right)^2\right],$$

$$Var(\widehat{CDR}_i) = C_{i,I-i}^2\left(\prod_{j=I-i}^{I-1}\hat{f}_j^2 + \underbrace{\mathbb{E}\left[(f_{I-i}^{b,I})^2\right]\prod_{j=I+1-i}^{I-1}\mathbb{E}\left[(f_j^{b,I+1})^2\right]}_{A1}\right.$$

$$\left. - 2\underbrace{\mathbb{E}\left[f_{I-i}^{b,I}\hat{f}_{I-i}\prod_{j=I+1-i}^{I-1}\hat{f}_j f_j^{b,I+1}\right]}_{B1}\right).$$

**Calculation of $A1$**

$$A1 = \left(\frac{\hat{\sigma}_{I-i}^2}{S_{I-i}^I} + \hat{f}_{I-i}^2\right)\prod_{j=I+1-i}^{I-1}\left[\left(\frac{C_{I-j,j}}{S_j^{I+1}}\right)^2\frac{\hat{\sigma}_j^2}{S_j^I} + \hat{f}_j^2\right],$$

$$A1 = \left(\prod_{j=I-i}^{I-1}\hat{f}_j^2\right)\left(1 + \frac{\hat{\sigma}_{I-i}^2}{\hat{f}_{I-i}^2 S_{I-i}^I}\right)\prod_{j=I+1-i}^{I-1}\left[\left(\frac{C_{I-j,j}}{S_j^{I+1}}\right)^2\frac{\hat{\sigma}_j^2}{\hat{f}_j^2 S_j^I} + 1\right],$$

where

$$\left(\frac{C_{I-j,j}}{S_j^{I+1}}\right)^2\frac{\hat{\sigma}_j^2}{\hat{f}_j^2 S_j^I} \leq \frac{\hat{\sigma}_j^2}{S_j^I} \approx 0.$$

Using the linear approximation

$$\prod_j(1+x_j) \approx 1 + \sum_j x_j,$$

we obtain

$$A1 \approx \left(\prod_{j=I-i}^{I-1}\hat{f}_j^2\right)\left[1 + \frac{\hat{\sigma}_{I-i}^2}{\hat{f}_{I-i}^2 S_{I-i}^I} + \sum_{j=I-i+1}^{I-1}\left(\frac{C_{I-j,j}}{S_j^{I+1}}\right)^2\frac{\hat{\sigma}_j^2}{\hat{f}_j^2 S_j^I}\right].$$

**Calculation of $B1$**

Using the independence of pseudo-development factors, we have

$$B1 = \prod_{j=I-i}^{I-1}\hat{f}_j^2.$$

Thus, we obtain

$$Var(\widehat{CDR}_i) \approx C_{i,I-i}^2\left[\prod_{j=I-i}^{I-1}\hat{f}_j^2 + \left(\prod_{j=I-i}^{I-1}\hat{f}_j^2\right)\left(1 + \frac{\hat{\sigma}_{I-i}^2}{\hat{f}_{I-i}^2 S_{I-i}^I} + \sum_{j=I-i+1}^{I-1}\left(\frac{C_{I-j,j}}{S_j^{I+1}}\right)^2\frac{\hat{\sigma}_j^2}{\hat{f}_j^2 S_j^I}\right) - 2\prod_{j=I-i}^{I-1}\hat{f}_j^2\right],$$

i.e.

$$Var(\widehat{CDR_i}) \approx \hat{C}_{i,I}^2 \left( \frac{\hat{\sigma}_{I-i}^2}{\hat{f}_{I-i}^2 S_{I-i}^I} + \sum_{j=I-i+1}^{I-1} \left( \frac{C_{I-j,j}}{S_j^{I+1}} \right)^2 \frac{\hat{\sigma}_j^2}{\hat{f}_j^2 S_j^I} \right). \tag{1}$$

This formula corresponds to the estimator of the estimation error proposed by Wüthrich *et al.* (2008) for accident year $i$ (equations (3.10) and (3.14)).

### *4.4.1.2. Estimation error for aggregated accident years*

The estimation error for aggregated accident years is written

$$Var\left( \sum_{i=1}^{I} \widehat{CDR_i} \right) = \sum_{i=1}^{I} Var(\widehat{CDR_i}) + 2 \sum_{1 \leq i < j \leq I} Cov(\widehat{CDR_i}, \widehat{CDR_j}),$$

with

$$Cov(\widehat{CDR_i}, \widehat{CDR_j}) = E(\widehat{CDR_i}\widehat{CDR_j}) - E(\widehat{CDR_i})E(\widehat{CDR_j}).$$

According to what precedes,

$$Cov(\widehat{CDR_i}, \widehat{CDR_j}) = E(\widehat{CDR_i}\widehat{CDR_j}).$$

With no process error, we can write $C_{i,I+1-i}^b = f_{I-i}^{b,I} C_{i,I-i}$. We suppose $i < j$, thus $I - i > I - j$. We have then

$\mathbb{E}(\widehat{CDR_i}\widehat{CDR_j})$

$$= \mathbb{E}\left[ \left( C_{i,I-i} \prod_{k=I-i}^{I-1} \hat{f}_k - f_{I-i}^{b,I} C_{i,I-i} \prod_{k=I+1-i}^{I-1} f_k^{b,I+1} \right) \left( C_{j,I-j} \prod_{k=I-j}^{I-1} \hat{f}_k - f_{I-j}^{b,I} C_{j,I-j} \prod_{k=I+1-j}^{I-1} f_k^{b,I+1} \right) \right]$$

$$= C_{i,I-i} C_{j,I-j} \mathbb{E}\left[ \left( \prod_{k=I-i}^{I-1} \hat{f}_k \right) \left( \prod_{k=I-j}^{I-1} \hat{f}_k \right) + \underbrace{f_{I-i}^{b,I} f_{I-j}^{b,I} \left( \prod_{k=I+1-i}^{I-1} f_k^{b,I+1} \right) \left( \prod_{k=I+1-j}^{I-1} f_k^{b,I+1} \right)}_{A} \right.$$

$$\left. - \underbrace{f_{I-i}^{b,I} \left( \prod_{k=I+1-i}^{I-1} f_k^{b,I+1} \right) \left( \prod_{k=I-j}^{I-1} \hat{f}_k \right)}_{B} - \underbrace{f_{I-j}^{b,I} \left( \prod_{k=I+1-j}^{I-1} f_k^{b,I+1} \right) \left( \prod_{k=I-i}^{I-1} \hat{f}_k \right)}_{C} \right].$$

Using the independence of pseudo-development factors seen at time $I + 1$, we obtain

$$\mathbb{E}[B] = \mathbb{E}[C] = \left( \prod_{k=I-i}^{I-1} \hat{f}_k \right) \left( \prod_{k=I-j}^{I-1} \hat{f}_k \right).$$

**Calculation of $\mathbb{E}[A]$**

$$\mathbb{E}[A] = \mathbb{E}\left[ f_{I-i}^{b,I} f_{I-j}^{b,I} \left( \prod_{k=I+1-i}^{I-1} f_k^{b,I+1} \right) \left( \prod_{k=I+1-j}^{I-1} f_k^{b,I+1} \right) \right],$$

and as $I - i > I - j$, we have

$$\mathbb{E}[A] = \mathbb{E}\left[ f_{I-j}^{b,I} \underbrace{f_{I-i}^{b,I} f_{I-i}^{b,I+1}}_{\text{not independents}} \left( \prod_{k=I+1-j}^{I-i-1} f_k^{b,I+1} \right) \left( \prod_{k=I+1-i}^{I-1} (f_k^{b,I+1})^2 \right) \right].$$

Using the independence of development factors for different years, we have

$$\mathbb{E}[A] = \underbrace{\mathbb{E}[f_{I-j}^{b,I}]}_{\hat{f}_{I-j}} \mathbb{E}[f_{I-i}^{b,I} f_{I-i}^{b,I+1}] \left( \prod_{k=I+1-j}^{I-i-1} \underbrace{\mathbb{E}[f_k^{b,I+1}]}_{\hat{f}_k} \right) \left( \prod_{k=I+1-i}^{I-1} \mathbb{E}\left[(f_k^{b,I+1})^2\right] \right).$$

Since
$$\mathbb{E}\left[(f_k^{b,I+1})^2\right] = \left(\frac{C_{I-k,k}}{S_k^{I+1}}\right)^2 \frac{\hat{\sigma}_k^2}{S_k^I} + \hat{f}_k^2,$$

and
$$f_{I-i}^{b,I+1} = \frac{S_{I-i}^I}{S_{I-i}^{I+1}} \hat{f}_{I-i} + f_{I-i}^{b,I} \frac{C_{i,I-i}}{S_{I-i}^{I+1}},$$

we write
$\mathbb{E}[A]$

$$= \hat{f}_{I-j} \mathbb{E}\left[ f_{I-i}^{b,I} \left( \frac{S_{I-i}^I}{S_{I-i}^{I+1}} \hat{f}_{I-i} + f_{I-i}^{b,I} \frac{C_{i,I-i}}{S_{I-i}^{I+1}} \right) \right] \left( \prod_{k=I+1-j}^{I-i-1} \hat{f}_k \right) \prod_{k=I+1-i}^{I-1} \left( \left(\frac{C_{I-k,k}}{S_k^{I+1}}\right)^2 \frac{\hat{\sigma}_k^2}{S_k^I} + \hat{f}_k^2 \right),$$

$$= \hat{f}_{I-j} \mathbb{E}\left[ f_{I-i}^{b,I} \left( \frac{S_{I-i}^I}{S_{I-i}^{I+1}} \hat{f}_{I-i} + f_{I-i}^{b,I} \frac{C_{i,I-i}}{S_{I-i}^{I+1}} \right) \right] \left( \prod_{k=I+1-j}^{I-i-1} \hat{f}_k \right) \left( \prod_{k=I+1-i}^{I-1} \hat{f}_k^2 \right) \prod_{k=I+1-i}^{I-1} \left( \left(\frac{C_{I-k,k}}{S_k^{I+1}}\right)^2 \frac{\hat{\sigma}_k^2}{S_k^I \hat{f}_k^2} + 1 \right),$$

$$= \hat{f}_{I-j} \left( \frac{S_{I-i}^I}{S_{I-i}^{I+1}} \hat{f}_{I-i} \mathbb{E}[f_{I-i}^{b,I}] \right.$$
$$\left. + \frac{C_{i,I-i}}{S_{I-i}^{I+1}} \mathbb{E}\left[(f_{I-i}^{b,I})^2\right] \right) \left( \prod_{k=I+1-j}^{I-i-1} \hat{f}_k \right) \left( \prod_{k=I+1-i}^{I-1} \hat{f}_k^2 \right) \prod_{k=I+1-i}^{I-1} \left( \left(\frac{C_{I-k,k}}{S_k^{I+1}}\right)^2 \frac{\hat{\sigma}_k^2}{S_k^I \hat{f}_k^2} + 1 \right).$$

We have $\mathbb{E}\left[(f_{I-i}^{b,I})^2\right] = \frac{\hat{\sigma}_{I-i}^2}{S_{I-i}^I} + \hat{f}_{I-i}^2$, thus we obtain

$$\mathbb{E}[A] = \hat{f}_{I-j} \left( \underbrace{\frac{S_{I-i}^I}{S_{I-i}^{I+1}} \hat{f}_{I-i}^2 + \frac{C_{i,I-i}}{S_{I-i}^{I+1}} \hat{f}_{I-i}^2}_{\hat{f}_{I-i}^2} + \frac{C_{i,I-i}}{S_{I-i}^{I+1}} \frac{\hat{\sigma}_{I-i}^2}{S_{I-i}^I} \right)$$
$$\times \left( \prod_{k=I+1-j}^{I-i-1} \hat{f}_k \right) \left( \prod_{k=I+1-i}^{I-1} \hat{f}_k^2 \right) \prod_{k=I+1-i}^{I-1} \left( \left(\frac{C_{I-k,k}}{S_k^{I+1}}\right)^2 \frac{\hat{\sigma}_k^2}{S_k^I \hat{f}_k^2} + 1 \right).$$

$\left(\frac{C_{I-k,k}}{S_k^{I+1}}\right)^2 \frac{\hat{\sigma}_k^2}{S_k^I \hat{f}_k^2} \leq \frac{\hat{\sigma}_k^2}{S_k^I} \approx 0$, so using the linear approximation

$$\prod_j (1 + x_j) \approx 1 + \sum_j x_j,$$

we have

$$\mathbb{E}[A] \approx \hat{f}_{I-j} \left( \hat{f}_{I-i}^2 + \frac{C_{i,I-i}}{S_{I-i}^{I+1}} \frac{\hat{\sigma}_{I-i}^2}{S_{I-i}^I} \right) \left( \prod_{k=I+1-j}^{I-i-1} \hat{f}_k \right) \left( \prod_{k=I+1-i}^{I-1} \hat{f}_k^2 \right) \left( 1 + \sum_{k=I+1-i}^{I-1} \left(\frac{C_{I-k,k}}{S_k^{I+1}}\right)^2 \frac{\hat{\sigma}_k^2}{S_k^I \hat{f}_k^2} \right),$$

$$\mathbb{E}[A] \approx \left( \prod_{k=I-i}^{I-1} \hat{f}_k \right) \left( \prod_{k=I-j}^{I-1} \hat{f}_k \right) \left( 1 + \underbrace{\frac{C_{i,I-i}}{S_{I-i}^{I+1}} \frac{\hat{\sigma}_{I-i}^2}{S_{I-i}^I \hat{f}_{I-i}^2}}_{\leq \frac{\hat{\sigma}_{I-i}^2}{S_{I-i}^I} \approx 0} \right) \left( 1 + \sum_{k=I+1-i}^{I-1} \underbrace{\left(\frac{C_{I-k,k}}{S_k^{I+1}}\right)^2 \frac{\hat{\sigma}_k^2}{S_k^I \hat{f}_k^2}}_{\leq \frac{\hat{\sigma}_k^2}{S_k^I} \approx 0} \right).$$

Using a similar linear approximation, we obtain

$$\mathbb{E}[A] \approx \left(\prod_{k=I-i}^{I-1} \hat{f}_k\right)\left(\prod_{k=I-j}^{I-1} \hat{f}_k\right)\left(1 + \frac{C_{i,I-i}}{S_{I-i}^{I+1}} \frac{\hat{\sigma}_{I-i}^2}{S_{I-i}^I \hat{f}_{I-i}^2} + \sum_{k=I+1-i}^{I-1} \left(\frac{C_{I-k,k}}{S_k^{I+1}}\right)^2 \frac{\hat{\sigma}_k^2}{S_k^I \hat{f}_k^2}\right).$$

Thus,

$$\mathbb{E}(\widehat{CDR}_i \widehat{CDR}_j) \approx C_{i,I} C_{j,I} \left(\frac{C_{i,I-i} \hat{\sigma}_{I-i}^2}{S_{I-i}^{I+1} \hat{f}_{I-i}^2 S_{I-i}^I} + \sum_{k=I-i+1}^{I-1} \left(\frac{C_{I-k,k}}{S_k^{I+1}}\right)^2 \frac{\hat{\sigma}_k^2}{\hat{f}_k^2 S_k^I}\right). \quad (2)$$

We find the same covariance as proposed by Wüthrich *et al.* (2008) (equation (3.13)), and lastly the same estimator $(u - bias)^2$ of the aggregated estimation error (equation (3.14)) thanks to equations (1) and (2).

### 4.4.2. Process error

#### 4.4.2.1. *Process error for single accident year*

By neglecting estimation error, we have $f_j^{b,I} = \hat{f}_j$.
Thus,

$$f_j^{b,I+1} = \frac{S_j^I}{S_j^{I+1}} \hat{f}_j + \frac{C_{I-j,j+1}^b}{S_j^{I+1}},$$

with $C_{I-j,j+1}^b = \hat{f}_j C_{I-j,j} + \hat{\sigma}_j \sqrt{C_{I-j,j}} \, \epsilon_{I-j,j+1}^b$ where $\epsilon_{I-j,j+1}^b \sim N(0,1)$.

The following results will be used thereafter, for $i \in \{0, \dots, I-1\}$ and $j \in \{0, \dots, I-1\}$:
- $Var(C_{i,I-i+1}^b) = C_{i,I-i} \hat{\sigma}_{I-i}^2$,
- $\mathbb{E}\left[(C_{i,I-i+1}^b)^2\right] = Var(C_{i,I-i+1}^b) + \mathbb{E}(C_{i,I-i+1}^b)^2 = C_{i,I-i} \hat{\sigma}_{I-i}^2 + C_{i,I-i}^2 \hat{f}_{I-i}^2$,
- $Var(f_j^{b,I+1}) = \frac{Var(C_{I-j,j+1}^b)}{\left(S_j^{I+1}\right)^2} = \frac{C_{I-j,j} \hat{\sigma}_j^2}{\left(S_j^{I+1}\right)^2}$,
- $\mathbb{E}\left[(f_j^{b,I+1})^2\right] = Var(f_j^{b,I+1}) + \mathbb{E}(f_j^{b,I+1})^2 = \frac{C_{I-j,j} \hat{\sigma}_j^2}{\left(S_j^{I+1}\right)^2} + \hat{f}_j^2.$

The variance of the CDR is written

$$Var(CDR_i) = Var\left(C_{i,I-i} \prod_{j=I-i}^{I-1} \hat{f}_j - C_{i,I-i+1}^b \prod_{j=I-i+1}^{I-1} f_j^{b,I+1}\right) = Var\left(C_{i,I-i+1}^b \prod_{j=I-i+1}^{I-1} f_j^{b,I+1}\right).$$

Using the independence of the $f_j^{b,I+1}$ for different development years, we have first

$$\mathbb{E}\left[\left(C_{i,I-i+1}^b \prod_{j=I-i+1}^{I-1} f_j^{b,I+1}\right)^2\right] = \mathbb{E}\left[(C_{i,I-i+1}^b)^2\right] \prod_{j=I-i+1}^{I-1} \mathbb{E}\left[(f_j^{b,I+1})^2\right],$$

then

$$\mathbb{E}\left[\left(C_{i,I-i+1}^b \prod_{j=I-i+1}^{I-1} f_j^{b,I+1}\right)^2\right] = (C_{i,I-i} \hat{\sigma}_{I-i}^2 + C_{i,I-i}^2 \hat{f}_{I-i}^2) \prod_{j=I-i+1}^{I-1} \left(\hat{f}_j^2 + \frac{\hat{\sigma}_j^2 C_{I-j,j}}{\left(S_j^{I+1}\right)^2}\right).$$

Second, we have

$$\mathbb{E}\left[C_{i,I-i+1}^b \prod_{j=I-i+1}^{I-1} f_j^{b,I+1}\right] = C_{i,I-i}\hat{f}_{I-i} \prod_{j=I-i+1}^{I-1} \hat{f}_j = \hat{C}_{i,I}.$$

Finally, we obtain

$$Var(CDR_i) = \left(C_{i,I-i}\,\hat{\sigma}_{I-i}^2 + C_{i,I-i}^2\,\hat{f}_{I-i}^2\right) \prod_{j=I-i+1}^{I-1} \left(\hat{f}_j^2 + \frac{\hat{\sigma}_j^2 C_{I-j,j}}{\left(S_j^{I+1}\right)^2}\right) - \hat{C}_{i,I}^2,$$

which can be rewritten as

$$Var(CDR_i) = \hat{C}_{i,I}^2 \left[\left(1 + \frac{\hat{\sigma}_{I-i}^2}{\hat{f}_{I-i}^2 C_{i,I-i}}\right) \prod_{j=I-i+1}^{I-1} \left(1 + \frac{\hat{\sigma}_j^2 C_{I-j,j}}{\hat{f}_j^2 \left(S_j^{I+1}\right)^2}\right) - 1\right]. \quad (3)$$

This result is the same as the estimator of process error for a single accident year $i$ of Wüthrich *et al.* (2008), equation (3.17).

### *4.4.2.2.  Process error for aggregated accident years*

Process error for all accident years is written

$$Var\left(\sum_{i=1}^{I} CDR_i\right) = \sum_{i=1}^{I} Var(CDR_i) + 2 \sum_{1 \le i < j \le I} Cov(CDR_i, CDR_j),$$

with

$$Cov(CDR_i, CDR_j) = \mathbb{E}(CDR_i CDR_j) - \mathbb{E}(CDR_i)\mathbb{E}(CDR_j).$$

Here, we do not take into account estimation error, thus we have for $j \in \{0, \ldots, I-1\}$,

$$f_j^{b,I} = \hat{f}_j,$$

and

$$f_j^{b,I+1} = \frac{S_j^I}{S_j^{I+1}} \hat{f}_j + \frac{C_{I-j,j+1}^b}{S_j^{I+1}},$$

with

$$C_{I-j,j+1}^b = \hat{f}_j C_{I-j,j} + \hat{\sigma}_j \sqrt{C_{I-j,j}}\, \epsilon_{I-j,j+1}^b,$$

where $\epsilon_{I-j,j+1}^b \sim N(0,1)$.

We have

$$\mathbb{E}[CDR_i] = \mathbb{E}\left[C_{i,I-i} \prod_{j=I-i}^{I-1} \hat{f}_j - C_{i,I-i+1}^b \prod_{j=I-i+1}^{I-1} f_j^{b,I+1}\right],$$

$$\mathbb{E}[CDR_i] = C_{i,I-i} \prod_{j=I-i}^{I-1} \hat{f}_j - \underbrace{\mathbb{E}[C_{i,I-i+1}^b]}_{\hat{f}_{I-i} C_{i,I-i}} \prod_{j=I-i+1}^{I-1} \underbrace{\mathbb{E}[f_j^{b,I+1}]}_{\hat{f}_j} = 0,$$

so

$$Cov(CDR_i, CDR_j) = E(CDR_i CDR_j).$$

We suppose $i < j$ therefore $I - i > I - j$. The covariance is written
$Cov(CDR_i, CDR_j)$

$$= \mathbb{E}\left[\left(C_{i,I-i}\prod_{k=I-i}^{I-1}\hat{f}_k - C_{i,I-i+1}^b\prod_{k=I-i+1}^{I-1}f_k^{b,I+1}\right)\left(C_{j,I-j}\prod_{k=I-j}^{I-1}\hat{f}_k - C_{j,I-j+1}^b\prod_{k=I-j+1}^{I-1}f_k^{b,I+1}\right)\right].$$

Using the independence of pseudo-development factors for several development years, we obtain

$Cov(CDR_i, CDR_j)$

$$= C_{i,I-i}C_{j,I-j}\left(\prod_{k=I-i}^{I-1}\hat{f}_k\right)\left(\prod_{k=I-j}^{I-1}\hat{f}_k\right) - C_{i,I-i}\left(\prod_{k=I-i}^{I-1}\hat{f}_k\right)\underbrace{\mathbb{E}[C_{j,I-j+1}^b]}_{\hat{f}_{I-j}C_{j,I-j}}\prod_{k=I-j+1}^{I-1}\underbrace{\mathbb{E}[f_k^{b,I+1}]}_{\hat{f}_k}$$

$$- C_{j,I-j}\left(\prod_{k=I-j}^{I-1}\hat{f}_k\right)\underbrace{\mathbb{E}[C_{i,I-i+1}^b]}_{\hat{f}_{I-i}C_{i,I-i}}\prod_{k=I-i+1}^{I-1}\underbrace{\mathbb{E}[f_k^{b,I+1}]}_{\hat{f}_k}$$

$$+ \mathbb{E}\left[\underbrace{C_{i,I-i+1}^b f_{I-i}^{b,I+1}}_{\text{not independents}} C_{j,I-j+1}^b \left(\prod_{k=I-i+1}^{I-1}(f_k^{b,I+1})^2\right)\prod_{k=I-j+1}^{I-i-1}f_k^{b,I+1}\right],$$

so we have

$\mathbb{E}[CDR_i CDR_j]$

$$= -\hat{C}_{i,I}\hat{C}_{j,I} + \underbrace{\mathbb{E}\left[C_{i,I-i+1}^b\left(\frac{S_{I-i}^I}{S_{I-i}^{I+1}}\hat{f}_{I-i} + \frac{C_{i,I-i+1}^b}{S_{I-i}^{I+1}}\right)\right]}_{D}\underbrace{\mathbb{E}[C_{j,I-j+1}^b]}_{\hat{f}_{I-j}C_{j,I-j}}\prod_{k=I-i+1}^{I-1}\underbrace{\mathbb{E}[(f_k^{b,I+1})^2]}_{\frac{\hat{\sigma}_k^2 C_{I-k,k}}{(S_k^{I+1})^2}+\hat{f}_k^2}\prod_{k=I-j+1}^{I-i-1}\underbrace{\mathbb{E}[f_k^{b,I+1}]}_{\hat{f}_k}.$$

**Calculation of D**

$$D = \frac{S_{I-i}^I}{S_{I-i}^{I+1}}\hat{f}_{I-i}\underbrace{\mathbb{E}[C_{i,I-i+1}^b]}_{\hat{f}_{I-i}C_{i,I-i}} + \underbrace{\mathbb{E}[(C_{i,I-i+1}^b)^2]}_{C_{i,I-i}\hat{\sigma}_{I-i}^2 + C_{i,I-i}^2\hat{f}_{I-i}^2}/S_{I-i}^{I+1} = \left(1 + \frac{\hat{\sigma}_{I-i}^2}{\hat{f}_{I-i}^2 S_{I-i}^{I+1}}\right)\hat{f}_{I-i}^2 C_{i,I-i},$$

so we have

$$\mathbb{E}[CDR_i CDR_j] = \left(1 + \frac{\hat{\sigma}_{I-i}^2}{\hat{f}_{I-i}^2 S_{I-i}^{I+1}}\right)\hat{f}_{I-i}^2 C_{i,I-i}\hat{f}_{I-j}C_{j,I-j}\left(\prod_{k=I-j+1}^{I-i-1}\hat{f}_k\right)\prod_{k=I-i+1}^{I-1}\left(\frac{\hat{\sigma}_k^2 C_{I-k,k}}{(S_k^{I+1})^2}+\hat{f}_k^2\right)$$
$$- \hat{C}_{i,I}\hat{C}_{j,I}.$$

Finally, we obtain the following result:

$$\boxed{\mathbb{E}[CDR_i CDR_j] = \hat{C}_{i,I}\hat{C}_{j,I}\left[\left(1 + \frac{\hat{\sigma}_{I-i}^2}{\hat{f}_{I-i}^2 S_{I-i}^{I+1}}\right)\prod_{k=I-i+1}^{I-1}\left(1 + \frac{\hat{\sigma}_k^2 C_{I-k,k}}{\hat{f}_k^2 (S_k^{I+1})^2}\right) - 1\right].} \quad (4)$$

This result corresponds to equation (3.18) of Wüthrich *et al.* (2008) and makes it possible to obtain the aggregate process error of equation (3.16), thanks to equations (3) and (4).

### 4.4.3. Prediction error

The calculation of the total variance amounts to summing estimation error and process error, those being orthogonal (see Appendix). The prediction error for accident year $i$ is thus
$$Var(CDR_i) + Var(\widehat{CDR_i}),$$
and the prediction error for aggregated accident years is
$$Var(CDR) + Var(\widehat{CDR}).$$

## 4.5. Closed-form expressions including a tail factor

In this section, we propose closed-form expressions for estimation, process and prediction errors including a tail factor, whose modeling is described in 4.2. Let $\widehat{CDR}_i^{TF}$ denote the CDR taking into account only estimation error and $CDR_i^{TF}$ the CDR taking into account pure process error, those two CDR including a tail factor.

### 4.5.1. Estimation error

Initially, we calculate the expected tail factor $\hat{f}_{ult}$ by the following linear extrapolation:
$$\forall j \in \{0, \ldots, I-1\}, \quad \ln(\hat{f}_j - 1) = \hat{a}.j + \hat{b},$$
with
$$\hat{f}_{ult} = \prod_{j=I}^{I_{ult}-1} \left(1 + e^{\hat{a}.j + \hat{b}}\right).$$

In each bootstrap iteration, in order to take into account the estimation error of the parameters $\hat{a}$ and $\hat{b}$, we simulate a tail factor $f_{ult}^b$ with mean $\hat{f}_{ult}$ and variance $\sigma_{ult}^2$.

#### 4.5.1.1. Estimation error for a single accident year

For $i \in \{1, \ldots, I\}$, the variance of the CDR is written
$$Var(\widehat{CDR}_i^{TF}) = Var\left[C_{i,I-i}\left(\prod_{j=I-i}^{I-1} \hat{f}_j\right)\hat{f}_{ult} - f_{I-i}^{b,I} C_{i,I-i}\left(\prod_{j=I+1-i}^{I-1} f_j^{b,I+1}\right) f_{ult}^b\right].$$

By the independent simulation of the tail factor $f_{ult}^b$, we obtain
$$\mathbb{E}\left[C_{i,I-i}\left(\prod_{j=I-i}^{I-1} \hat{f}_j\right)\hat{f}_{ult} - f_{I-i}^{b,I} C_{i,I-i}\left(\prod_{j=I+1-i}^{I-1} f_j^{b,I+1}\right) f_{ult}^b\right]$$
$$= C_{i,I-i}\left(\prod_{j=I-i}^{I-1} \hat{f}_j\right)\hat{f}_{ult} - \hat{f}_{I-i} C_{i,I-i}\left(\prod_{j=I-i+1}^{I-1} \hat{f}_j\right) \underbrace{\mathbb{E}[f_{ult}^b]}_{\hat{f}_{ult}}$$
$$= 0.$$

Thus, the variance is
$$Var(\widehat{CDR}_i^{TF}) = \mathbb{E}\left[\left(C_{i,I-i}\left(\prod_{j=I-i}^{I-1} \hat{f}_j\right)\hat{f}_{ult} - f_{I-i}^{b,I} C_{i,I-i}\left(\prod_{j=I+1-i}^{I-1} f_j^{b,I+1}\right) f_{ult}^b\right)^2\right].$$

Still by the independence argument, this expression can be rewritten as

$$Var(\widehat{CDR}_i^{TF}) = C_{i,I-i}^2 \left( \left( \prod_{j=I-i}^{I-1} \hat{f}_j^2 \right) \hat{f}_{ult}^2 + \underbrace{\mathbb{E}\left[(f_{I-i}^{b,I})^2\right] \prod_{j=I+1-i}^{I-1} \mathbb{E}\left[(f_j^{b,I+1})^2\right]}_{A1} \underbrace{\mathbb{E}\left[(f_{ult}^b)^2\right]}_{\sigma_{ult}^2 + \hat{f}_{ult}^2} \right.$$

$$\left. - 2 \underbrace{\mathbb{E}\left[f_{I-i}^{b,I} \hat{f}_{I-i} \prod_{j=I+1-i}^{I-1} \hat{f}_j f_j^{b,I+1}\right]}_{B1} \underbrace{\hat{f}_{ult} \mathbb{E}[f_{ult}^b]}_{\hat{f}_{ult}^2} \right).$$

Using the formulas leading to equation (1), we obtain

$$Var(\widehat{CDR}_i^{TF}) = C_{i,I-i}^2 \left( \left( \prod_{j=I-i}^{I-1} \hat{f}_j^2 \right) \hat{f}_{ult}^2 \right.$$

$$+ \left( \prod_{j=I-i}^{I-1} \hat{f}_j^2 \right) \left[ 1 + \frac{\hat{\sigma}_{I-i}^2}{\hat{f}_{I-i}^2 S_{I-i}^I} + \sum_{j=I-i+1}^{I-1} \left( \frac{C_{I-j,j}}{S_j^{I+1}} \right)^2 \frac{\hat{\sigma}_j^2}{\hat{f}_j^2 S_j^I} \right] (\sigma_{ult}^2 + \hat{f}_{ult}^2)$$

$$\left. - 2 \left( \prod_{j=I-i}^{I-1} \hat{f}_j^2 \right) \hat{f}_{ult}^2 \right),$$

i.e.

$$Var(\widehat{CDR}_i^{TF}) = \hat{C}_{i,I}^2 \left[ (\sigma_{ult}^2 + \hat{f}_{ult}^2) \left( 1 + \frac{\hat{\sigma}_{I-i}^2}{\hat{f}_{I-i}^2 S_{I-i}^I} + \sum_{j=I-i+1}^{I-1} \left( \frac{C_{I-j,j}}{S_j^{I+1}} \right)^2 \frac{\hat{\sigma}_j^2}{\hat{f}_j^2 S_j^I} \right) - \hat{f}_{ult}^2 \right].$$

Finally, the variance is

$$Var(\widehat{CDR}_i^{TF}) = \hat{C}_{i,I_{ult}}^2 \left[ \left( 1 + \frac{\sigma_{ult}^2}{\hat{f}_{ult}^2} \right) \left( 1 + \frac{\hat{\sigma}_{I-i}^2}{\hat{f}_{I-i}^2 S_{I-i}^I} + \sum_{j=I-i+1}^{I-1} \left( \frac{C_{I-j,j}}{S_j^{I+1}} \right)^2 \frac{\hat{\sigma}_j^2}{\hat{f}_j^2 S_j^I} \right) - 1 \right]. \quad (5a)$$

This result corresponds to the process error of the simulated distribution, including a tail factor and for accident year $i \geq 1$.

For accident year 0, we have

$$Var(\widehat{CDR}_0^{TF}) = Var(C_{0,I} \hat{f}_{ult} - C_{0,I} f_{ult}^b),$$

thus

$$Var(\widehat{CDR}_0^{TF}) = Var(C_{0,I} f_{ult}^b) = C_{0,I}^2 \sigma_{ult}^2. \quad (5b)$$

#### 4.5.1.2. Estimation error for aggregated accident years

With no process variance, we have $C_{i,I+1-i}^b = f_{I-i}^{b,I} C_{i,I-i}$. Let $i \in \{1, \dots, I\}$ and $j \in \{i+1, \dots, I\}$.
We have $I - i > I - j$, and

$$Cov(\widehat{CDR}_i^{TF}, \widehat{CDR}_j^{TF}) = \mathbb{E}(\widehat{CDR}_i^{TF} \widehat{CDR}_j^{TF}) - \underbrace{\mathbb{E}(\widehat{CDR}_i^{TF})}_{0} \underbrace{\mathbb{E}(\widehat{CDR}_j^{TF})}_{0}.$$

We have

$$Cov(\widehat{CDR}_i^{TF}, \widehat{CDR}_j^{TF})$$

$$= \mathbb{E}\left[\left(C_{i,I-i}\left(\prod_{k=I-i}^{I-1} \hat{f}_k\right)\hat{f}_{ult} - f_{I-i}^{b,I}C_{i,I-i}\left(\prod_{k=I+1-i}^{I-1} f_k^{b,I+1}\right)f_{ult}^b\right)\right.$$

$$\left.\times\left(C_{j,I-j}\left(\prod_{k=I-j}^{I-1} \hat{f}_k\right)\hat{f}_{ult} - f_{I-j}^{b,I}C_{j,I-j}\left(\prod_{k=I+1-j}^{I-1} f_k^{b,I+1}\right)f_{ult}^b\right)\right],$$

and using the independence properties, we obtain
$Cov(\widehat{CDR}_i^{TF}, \widehat{CDR}_j^{TF})$

$$= C_{i,I-i}C_{j,I-j}\left\{\left(\prod_{k=I-i}^{I-1} \hat{f}_k\right)\left(\prod_{k=I-j}^{I-1} \hat{f}_k\right)\hat{f}_{ult}^2 + \mathbb{E}\left[(f_{ult}^b)^2\right]\mathbb{E}[A] - \hat{f}_{ult}\mathbb{E}[f_{ult}^b]\mathbb{E}[B] - \hat{f}_{ult}\mathbb{E}[f_{ult}^b]\mathbb{E}[C]\right\}.$$

Using the results leading to equation (2), we obtain

$$\mathbb{E}[A] \approx \left(\prod_{k=I-i}^{I-1} \hat{f}_k\right)\left(\prod_{k=I-j}^{I-1} \hat{f}_k\right)\left(1 + \frac{C_{i,I-i}}{S_{I-i}^{I+1}}\frac{\hat{\sigma}_{I-i}^2}{S_{I-i}^I \hat{f}_{I-i}^2} + \sum_{k=I+1-i}^{I-1}\left(\frac{C_{I-k,k}}{S_k^{I+1}}\right)^2\frac{\hat{\sigma}_k^2}{S_k^I \hat{f}_k^2}\right),$$

and

$$\mathbb{E}[B] = \mathbb{E}[C] = \left(\prod_{k=I-i}^{I-1} \hat{f}_k\right)\left(\prod_{k=I-j}^{I-1} \hat{f}_k\right),$$

thus
$Cov(\widehat{CDR}_i^{TF}, \widehat{CDR}_j^{TF})$

$$\approx \hat{C}_{i,I}\hat{C}_{j,I}\left[(\sigma_{ult}^2 + \hat{f}_{ult}^2)\left(1 + \frac{C_{i,I-i}}{S_{I-i}^{I+1}}\frac{\hat{\sigma}_{I-i}^2}{S_{I-i}^I \hat{f}_{I-i}^2} + \sum_{k=I+1-i}^{I-1}\left(\frac{C_{I-k,k}}{S_k^{I+1}}\right)^2\frac{\hat{\sigma}_k^2}{S_k^I \hat{f}_k^2}\right) - \hat{f}_{ult}^2\right].$$

Finally, we obtain the covariance including estimation error by
$Cov(\widehat{CDR}_i^{TF}, \widehat{CDR}_j^{TF})$

$$\approx \hat{C}_{i,I_{ult}}\hat{C}_{j,I_{ult}}\left[\left(1 + \frac{\sigma_{ult}^2}{\hat{f}_{ult}^2}\right)\left(1 + \frac{C_{i,I-i}}{S_{I-i}^{I+1}}\frac{\hat{\sigma}_{I-i}^2}{S_{I-i}^I \hat{f}_{I-i}^2} + \sum_{k=I+1-i}^{I-1}\left(\frac{C_{I-k,k}}{S_k^{I+1}}\right)^2\frac{\hat{\sigma}_k^2}{S_k^I \hat{f}_k^2}\right) - 1\right] \quad (6a)$$

In addition, for $i \geq 1$ we have
$Cov(\widehat{CDR}_i^{TF}, \widehat{CDR}_0^{TF})$

$$= \mathbb{E}\left[\left(C_{i,I-i}\left(\prod_{k=I-i}^{I-1} \hat{f}_k\right)\hat{f}_{ult} - f_{I-i}^{b,I}C_{i,I-i}\left(\prod_{k=I+1-i}^{I-1} f_k^{b,I+1}\right)f_{ult}^b\right)(C_{0,I}\hat{f}_{ult} - C_{0,I}f_{ult}^b)\right].$$

Using the independence properties, we obtain

$$Cov(\widehat{CDR}_i^{TF}, \widehat{CDR}_0^{TF}) = -\hat{C}_{i,I_{ult}}\hat{C}_{0,I_{ult}} + C_{0,I}C_{i,I-i}\mathbb{E}[f_{I-i}^{b,I}]\left(\prod_{k=I+1-i}^{I-1} \mathbb{E}[f_k^{b,I+1}]\right)\mathbb{E}\left[(f_{ult}^b)^2\right],$$

i.e.

$$Cov(\widehat{CDR}_i^{TF}, \widehat{CDR}_0^{TF}) = \hat{C}_{i,I_{ult}}\hat{C}_{0,I_{ult}}\frac{\sigma_{ult}^2}{\hat{f}_{ult}^2}. \quad (6b)$$

Lastly, equations (5) and (6) allow to calculate the estimation error for aggregated accident years, including a tail factor, by

$$Var\left(\sum_{i=0}^{I}\widehat{CDR}_i^{TF}\right) = \sum_{i=0}^{I} Var(\widehat{CDR}_i^{TF}) + 2\sum_{0\le i<j\le I} Cov(\widehat{CDR}_i^{TF},\widehat{CDR}_j^{TF}). \quad (7)$$

### 4.5.2. Process error

With no estimation error, the tail factor is seen as an expected value: $f_{ult}^b = \hat{f}_{ult}$.

#### 4.5.2.1. Process error for a single accident year

The simulated CDR is written, for $i \in \{1, \dots, I\}$,

$$CDR_i^{TF} = C_{i,I-i}\left(\prod_{j=I-i}^{I-1}\hat{f}_j\right)\hat{f}_{ult} - C_{i,I-i+1}^b\left(\prod_{j=I-i+1}^{I-1} f_j^{b,I+1}\right)\hat{f}_{ult}.$$

Thus

$$Var(CDR_i^{TF}) = \hat{f}_{ult}^2\, Var\left(C_{i,I-i}\prod_{j=I-i}^{I-1}\hat{f}_j - C_{i,I-i+1}^b \prod_{j=I-i+1}^{I-1} f_j^{b,I+1}\right),$$

and using equation (3),

$$Var(CDR_i^{TF}) = \hat{f}_{ult}^2 \hat{C}_{i,I}^2 \left[\left(1+\frac{\hat{\sigma}_{I-i}^2}{\hat{f}_{I-i}^2 C_{i,I-i}}\right)\prod_{j=I-i+1}^{I-1}\left(1+\frac{\hat{\sigma}_j^2 C_{I-j,j}}{\hat{f}_j^2 (S_j^{I+1})^2}\right) - 1\right].$$

Process error for accident year $i \ge 1$ is thus written

$$\boxed{Var(CDR_i^{TF}) = \hat{C}_{i,I_{ult}}^2 \left[\left(1+\frac{\hat{\sigma}_{I-i}^2}{\hat{f}_{I-i}^2 C_{i,I-i}}\right)\prod_{j=I-i+1}^{I-1}\left(1+\frac{\hat{\sigma}_j^2 C_{I-j,j}}{\hat{f}_j^2 (S_j^{I+1})^2}\right) - 1\right].} \quad (8a)$$

We also have

$$\boxed{Var(CDR_0^{TF}) = Var(C_{0,I}\hat{f}_{ult} - C_{0,I}\hat{f}_{ult}) = 0.} \quad (8b)$$

#### 4.5.2.2. Process error for aggregated accident years

Let $i \in \{1,\dots,I\}$ and $j \in \{i+1,\dots,I\}$. The covariance is written

$$Cov(CDR_i^{TF}, CDR_j^{TF}) = \mathbb{E}[CDR_i^{TF} CDR_j^{TF}] - \underbrace{\mathbb{E}[CDR_i^{TF}]}_{0}\underbrace{\mathbb{E}[CDR_j^{TF}]}_{0} = \hat{f}_{ult}^2\, \mathbb{E}[CDR_i CDR_j].$$

Using equation (4), we obtain

$$\mathbb{E}[CDR_i^{TF} CDR_j^{TF}] = \hat{f}_{ult}^2 \hat{C}_{i,I}\hat{C}_{j,I}\left[\left(1+\frac{\hat{\sigma}_{I-i}^2}{\hat{f}_{I-i}^2 S_{I-i}^{I+1}}\right)\prod_{k=I-i+1}^{I-1}\left(1+\frac{\hat{\sigma}_k^2 C_{I-k,k}}{\hat{f}_k^2 (S_k^{I+1})^2}\right) - 1\right],$$

which can be written as

$$\boxed{\mathbb{E}[CDR_i^{TF} CDR_j^{TF}] = \hat{C}_{i,I_{ult}}\hat{C}_{j,I_{ult}}\left[\left(1+\frac{\hat{\sigma}_{I-i}^2}{\hat{f}_{I-i}^2 S_{I-i}^{I+1}}\right)\prod_{k=I-i+1}^{I-1}\left(1+\frac{\hat{\sigma}_k^2 C_{I-k,k}}{\hat{f}_k^2 (S_k^{I+1})^2}\right) - 1\right].} \quad (9a)$$

We also have, for $i \geq 1$,

$$\mathbb{E}\left[CDR_i^{TF} \underbrace{CDR_0^{TF}}_{0}\right] = 0. \quad (9b)$$

Equations (8) and (9) allow to calculate the process error including a tail factor for aggregated accident years by

$$Var\left(\sum_{i=0}^{I} CDR_i^{TF}\right) = \sum_{i=0}^{I} Var(CDR_i^{TF}) + 2 \sum_{0 \leq i < j \leq I} Cov(CDR_i^{TF}, CDR_j^{TF}). \quad (10)$$

### 4.5.3. Prediction error

According to the results in Appendix which can be extended to the inclusion of a tail factor, process error and estimation error are orthogonal. Thus, the prediction error for a single accident year $i$ can be written as

$$Var(CDR_i^{TF}) + Var(\widehat{CDR}_i^{TF}). \quad (11)$$

The prediction error for aggregated accident years is given by

$$Var\left(\sum_{i=0}^{I} CDR_i^{TF}\right) + Var\left(\sum_{i=0}^{I} \widehat{CDR}_i^{TF}\right). \quad (12)$$

## 4.6. Numerical example

We apply the bootstrap method described in 4.3.1 and the previous closed-form expressions to the loss development triangle used by Wüthrich *et al.* (2008). The results are obtained with 300 000 simulations, and we use the adaptations proposed in 4.3.2 for the calculation of the two kinds of error. The errors calculated thereafter refer to the standard deviation of the simulated empirical distribution. Moreover, the expressions leading to the analytical results are pointed out, and calculated by taking their square root to compare results homogeneous to first order moments.

We study the case where the development of the triangle is complete at time $I = 8$, as well as the inclusion of a tail factor to obtain cumulative payments at time $I_{ult} = 10$ in this example. Table 1 below presents the Chain Ladder factors and the tail development factor, and also the corresponding volatilities.

| Development year $j$ | 0 | 1 | 2 | 3 | 4 | 5 | 6 | 7 | ultimate |
|---|---|---|---|---|---|---|---|---|---|
| $\hat{f}_j / \hat{f}_{ult}$ | 1.47593 | 1.07190 | 1.02315 | 1.01613 | 1.00629 | 1.00559 | 1.00127 | 1.00112 | 1.00049 |
| $\hat{\sigma}_j^2 / \sigma_{ult}^2$ | 911.44 | 189.82 | 97.82 | 178.75 | 20.64 | 3.23 | 0.36 | 0.04 | 3.17E-08[2] |

**Table 1:** Development factors and corresponding volatilities.

---

[2] The variance of the tail factor is calculated by expression (*) of the Delta method, shown in 4.2.

### 4.6.1. Numerical results without a tail factor

Table 2 below shows the numerical results of the various errors without a tail factor.

| Accident year $i$ | Prediction error | | | Estimation error | | | Process error | | |
|---|---|---|---|---|---|---|---|---|---|
| | Analytical[3] | Simulated | Relative distance[4] | Analytical[5] | Simulated | Relative distance4 | Analytical[6] | Simulated | Relative distance4 |
| 0 | - | - | - | - | - | - | - | - | - |
| 1 | 566 | 567 | 0.14% | 406 | 406 | 0.07% | 394 | 394 | 0.03% |
| 2 | 1 487 | 1 486 | 0.05% | 875 | 874 | 0.14% | 1 201 | 1 201 | 0.03% |
| 3 | 3 923 | 3 918 | 0.13% | 1 922 | 1 923 | 0.05% | 3 420 | 3 427 | 0.22% |
| 4 | 9 723 | 9 707 | 0.16% | 4 298 | 4 304 | 0.14% | 8 721 | 8 690 | 0.36% |
| 5 | 28 443 | 28 411 | 0.11% | 11 636 | 11 644 | 0.07% | 25 953 | 25 947 | 0.02% |
| 6 | 20 954 | 20 974 | 0.09% | 7 863 | 7 863 | 0.01% | 19 423 | 19 423 | 0.00% |
| 7 | 28 119 | 28 148 | 0.10% | 9 836 | 9 823 | 0.13% | 26 343 | 26 356 | 0.05% |
| 8 | 53 321 | 53 279 | 0.08% | 17 558 | 17 574 | 0.09% | 50 347 | 50 211 | 0.27% |
| Total | 81 081 | 81 074 | 0.01% | 29 784 | 29 795 | 0.04% | 75 412 | 75 433 | 0.03% |

**Table 2:** Prediction, estimation and process errors without a tail factor.

The relative distances for the various errors are very low which ensure here the replication of the estimators proposed by Wüthrich *et al.* (2008). These results also allow to check the linear approximations used in 4.4.

### 4.6.2. Numerical results including a tail factor

The procedure presented in 4.3.1 allows to simulate a tail factor independently in each bootstrap iteration (**step 5.b**) by a normal distribution with mean $\hat{f}_{ult} \approx 1,00049$ and variance $\sigma^2_{ult} = 3,17\text{E} - 08$. The calculation of these parameters is presented in 4.2. Table 3 below summarizes the numerical results including a tail factor:

| Accident year $i$ | Prediction error | | | Estimation error | | | Process error | | |
|---|---|---|---|---|---|---|---|---|---|
| | Analytical[7] | Simulated | Relative distance[8] | Analytical[9] | Simulated | Relative distance8 | Analytical[10] | Simulated | Relative distance8 |
| 0 | 655 | 655 | 0.00% | 655 | 655 | 0.00% | 0 | 0 | 0.00% |
| 1 | 897 | 896 | 0.11% | 806 | 805 | 0.12% | 394 | 393 | 0.26% |
| 2 | 1 642 | 1 641 | 0.06% | 1 119 | 1 118 | 0.09% | 1 202 | 1 199 | 0.24% |
| 3 | 3 976 | 3 975 | 0.03% | 2 026 | 2 027 | 0.05% | 3 422 | 3 425 | 0.11% |

---

[3] The prediction error is calculated starting from equations (3.10) and (3.17) for a single accident year and (3.15) for aggregated accident years.
[4] We present here absolute values of relative distances.
[5] The estimation error is calculated by means of equations (3.10) and (3.14).
[6] The process error is calculated by means of equations (3.17) and (3.16).
[7] The prediction error is obtained starting from equation (11) for a single accident year and equation (12) for aggregated accident years.
[8] We present here absolute values of relative distances.
[9] The estimation error is calculated by equations (5), (6) and (7).
[10] The process error is calculated by equations (8), (9) and (10).

| | | | | | | | | | |
|---|---|---|---|---|---|---|---|---|---|
| 4 | 9 749 | 9 739 | 0.10% | 4 349 | 4 348 | 0.02% | 8 726 | 8 739 | 0.15% |
| 5 | 28 464 | 28 479 | 0.05% | 11 661 | 11 641 | 0.17% | 25 966 | 25 942 | 0.09% |
| 6 | 20 974 | 20 959 | 0.07% | 7 893 | 7 890 | 0.04% | 19 433 | 19 397 | 0.18% |
| 7 | 28 140 | 28 130 | 0.04% | 9 861 | 9 849 | 0.12% | 26 356 | 26 340 | 0.06% |
| 8 | 53 351 | 53 320 | 0.06% | 17 578 | 17 554 | 0.14% | 50 372 | 50 409 | 0.07% |
| Total | 81 336 | 81 249 | 0.11% | 30 381 | 30 326 | 0.18% | 75 449 | 75 465 | 0.02% |

**Table 3:** Prediction, estimation and process errors including a tail factor.

The measure of the standard deviation of the empirical distribution allows to check the closed-form expressions detailed in 4.5. We can notice that the inclusion of a tail factor increases the variability of the CDR at several levels:

- The estimation error is increased because of the direct effect of the volatility of the tail factor in the total variance (see equation (5)),
- The process error of the cumulative payments of the first sub-diagonal is diffused by best estimate calculation until time $I_{ult} > I$. The effect of the additional development year after time $I$ is to increase the variance of the CDR taking into account process variance (see equation (8)).

Finally, we obtain a higher prediction error when we include a tail factor, compared to the case of a complete development of the triangle at time $I$.

# Conclusion

An alternative method to the model of Wüthrich *et al.* (2008) has been developed in this study. This method consists in an adaptation of the « standard » bootstrap procedure and allows to measure the volatility in claims reserves over a one-year period, and to replicates the results of Wüthrich *et al.* (2008).

Compared to the model of Wüthrich *et al.* (2008), the adaptation of the bootstrap procedure has many advantages. In particular, this model provides a split between payments that will be done during the next year and best estimate calculation of claims reserves in one year. Therefore, the inclusion of this method in an internal model taking into account other risks is easy. The model developed in this study also includes a stochastic modeling of the tail factor: its use is therefore not restricted to loss triangles that are fully developed

This method can also be extended to measure the variability of the CDR in $K$ years, which is particularly useful within the ORSA framework. One will be able to obtain payments that will be done during the period $(I + K, I + K + 1]$ on one hand, and best estimate of claims reserves at time $I + K + 1$ on the other hand.

Several axes could supplement this study. We limited ourselves to a "stand-alone" vision, without taking account of a possible diversification effect between lines of business. Nevertheless, many insurance companies are led to model the dependencies between lines of business within their internal model. One of the axes of development would be to propose methods generalizing the bootstrap approach on the residuals in situation of dependencies. The correlation between the residuals of each triangle corresponding to the various lines of business could be modeled for example by means of copulas. Lastly, within the framework of an internal model, the reserve risk could be modeled jointly with other risks. It would then be necessary in this case to identify the links with the other risks (the premium risk in particular) and to integrate them in the model.

# Appendix

## Prediction error for a single accident year $i$

In this section, we want to decompose the prediction error for a single accident year $i$ into estimation error and process error. Let $\widehat{CDR}_i^b$ denote the CDR simulated by the bootstrap procedure detailed in 4.3.1. For accident year $i \geq 1$, the variance is written

$$Var(\widehat{CDR}_i^b) = Var\left(C_{i,I-i}\prod_{j=I-i}^{I-1}\hat{f}_j - C_{i,I-i+1}^b\prod_{j=I-i+1}^{I-1}f_j^{b,I+1}\right) = Var\left(C_{i,I-i+1}^b\prod_{j=I-i+1}^{I-1}f_j^{b,I+1}\right),$$

with
$C_{i,I-i+1}^b = f_{I-i}^{b,I}C_{i,I-i} + \hat{\sigma}_{I-i}\sqrt{C_{i,I-i}}\,\epsilon_{I-i}^b$ where $\epsilon_{I-i}^b \sim N(0,1)$,

and for all $j \in \{0,\ldots,I-1\}$, $f_j^{b,I} = \frac{\sum_{i=0}^{I-j-1}C_{i,j}f_{i,j}^{b,I}}{\sum_{i=0}^{I-j-1}C_{i,j}}$ where $f_{i,j}^{b,I} = r_{i,j}^b\sqrt{\frac{\hat{\sigma}_j^2}{C_{i,j}}} + \hat{f}_j$.

**Remark:** Random variables $\epsilon_j^b$ simulated to include process variance in the sub-diagonal are independent. Thus, those random variables are subscripted over the development years in order to reduce the notations, which is enough to differentiate these random variables.

We also have

$$\forall j \in \{0,\ldots,I-1\}, \quad f_j^{b,I+1} = \frac{\sum_{i=0}^{I-j-1}C_{i,j}f_{i,j} + C_{I-j,j+1}^b}{\sum_{i=0}^{I-j}C_{i,j}} = \frac{S_j^I}{S_j^{I+1}}\hat{f}_j + \frac{f_j^{b,I}C_{I-j,j}}{S_j^{I+1}} + \frac{\hat{\sigma}_j\sqrt{C_{I-j,j}}\epsilon_j^b}{S_j^{I+1}}.$$

Thus,

$$Var(\widehat{CDR}_i^b) = Var\left((f_{I-i}^{b,I}C_{i,I-i} + \hat{\sigma}_{I-i}\sqrt{C_{i,I-i}}\,\epsilon_{I-i}^b)\prod_{j=I+1-i}^{I-1}\left(\frac{S_j^I}{S_j^{I+1}}\hat{f}_j + \frac{f_j^{b,I}C_{I-j,j}}{S_j^{I+1}} + \frac{\hat{\sigma}_j\sqrt{C_{I-j,j}}\epsilon_j^b}{S_j^{I+1}}\right)\right).$$

We will use thereafter the following independence properties:
- $f_j^{b,I}$ and $f_k^{b,I}$ are independent for $j \neq k$,
- $\epsilon_j^b$ and $\epsilon_k^b$ are independent for $j \neq k$,
- $\epsilon_j^b$ and $f_k^{b,I}$ are independent for all $j, k$.

Thus,

$$\mathbb{E}\left[(f_{I-i}^{b,I}C_{i,I-i} + \hat{\sigma}_{I-i}\sqrt{C_{i,I-i}}\,\epsilon_{I-i}^b)\prod_{j=I+1-i}^{I-1}\left(\frac{S_j^I}{S_j^{I+1}}\hat{f}_j + \frac{f_j^{b,I}C_{I-j,j}}{S_j^{I+1}} + \frac{\hat{\sigma}_j\sqrt{C_{I-j,j}}\epsilon_j^b}{S_j^{I+1}}\right)\right]$$

$$= \underbrace{\mathbb{E}[f_{I-i}^{b,I}C_{i,I-i} + \hat{\sigma}_{I-i}\sqrt{C_{i,I-i}}\,\epsilon_{I-i}^b]}_{\hat{f}_{I-i}C_{i,I-i}}\prod_{j=I+1-i}^{I-1}\underbrace{\mathbb{E}\left[\frac{S_j^I}{S_j^{I+1}}\hat{f}_j + \frac{f_j^{b,I}C_{I-j,j}}{S_j^{I+1}} + \frac{\hat{\sigma}_j\sqrt{C_{I-j,j}}\epsilon_j^b}{S_j^{I+1}}\right]}_{\hat{f}_j}$$

$$= \hat{C}_{i,I}.$$

We also have

$$\mathbb{E}\left[\left((f_{I-i}^{b,I}C_{i,I-i} + \hat{\sigma}_{I-i}\sqrt{C_{i,I-i}}\,\epsilon_{I-i}^b)\prod_{j=I+1-i}^{I-1}\left(\frac{S_j^I}{S_j^{I+1}}\hat{f}_j + \frac{f_j^{b,I}C_{I-j,j}}{S_j^{I+1}} + \frac{\hat{\sigma}_j\sqrt{C_{I-j,j}}\epsilon_j^b}{S_j^{I+1}}\right)\right)^2\right]$$

$$= \underbrace{\mathbb{E}\left[(f_{I-i}^{b,I}C_{i,I-i} + \hat{\sigma}_{I-i}\sqrt{C_{i,I-i}}\,\epsilon_{I-i}^b)^2\right]}_{\alpha_i} \prod_{j=I+1-i}^{I-1} \underbrace{\mathbb{E}\left[\left(\frac{S_j^I}{S_j^{I+1}}\hat{f}_j + \frac{f_j^{b,I}C_{I-j,j}}{S_j^{I+1}} + \frac{\hat{\sigma}_j\sqrt{C_{I-j,j}}\epsilon_j^b}{S_j^{I+1}}\right)^2\right]}_{\beta_j}.$$

**Calculation of $\alpha_i$**

$$\alpha_i = \mathbb{E}\left[(f_{I-i}^{b,I}C_{i,I-i})^2\right] + \mathbb{E}\left[(\hat{\sigma}_{I-i}\sqrt{C_{i,I-i}}\,\epsilon_{I-i}^b)^2\right] + 2\mathbb{E}[f_{I-i}^{b,I}C_{i,I-i}]\underbrace{\mathbb{E}[\hat{\sigma}_{I-i}\sqrt{C_{i,I-i}}\,\epsilon_{I-i}^b]}_{0}.$$

Using results detailed in 4.4.1.1, we obtain

$$\alpha_i = C_{i,I-i}^2\left(\frac{\hat{\sigma}_{I-i}^2}{S_{I-i}^I} + \hat{f}_{I-i}^2\right) + \hat{\sigma}_{I-i}^2 C_{i,I-i}.$$

**Calculation of $\beta_j$**

$$\beta_j = Var\left(\frac{S_j^I}{S_j^{I+1}}\hat{f}_j + \frac{f_j^{b,I}C_{I-j,j}}{S_j^{I+1}} + \frac{\hat{\sigma}_j\sqrt{C_{I-j,j}}\epsilon_j^b}{S_j^{I+1}}\right) + \mathbb{E}\left[\frac{S_j^I}{S_j^{I+1}}\hat{f}_j + \frac{f_j^{b,I}C_{I-j,j}}{S_j^{I+1}} + \frac{\hat{\sigma}_j\sqrt{C_{I-j,j}}\epsilon_j^b}{S_j^{I+1}}\right]^2,$$

so we have

$$\beta_j = Var\left(\frac{f_j^{b,I}C_{I-j,j}}{S_j^{I+1}}\right) + Var\left(\frac{\hat{\sigma}_j\sqrt{C_{I-j,j}}\epsilon_j^b}{S_j^{I+1}}\right) + \hat{f}_j^2,$$

then

$$\beta_j = \left(\frac{C_{I-j,j}}{S_j^{I+1}}\right)^2 \frac{\hat{\sigma}_j^2}{S_j^I} + \frac{\hat{\sigma}_j^2 C_{I-j,j}}{(S_j^{I+1})^2} + \hat{f}_j^2.$$

We thus have

$$Var(\widehat{CDR}_i^b) = \left(C_{i,I-i}^2\left(\frac{\hat{\sigma}_{I-i}^2}{S_{I-i}^I} + \hat{f}_{I-i}^2\right) + \hat{\sigma}_{I-i}^2 C_{i,I-i}\right) \prod_{j=I+1-i}^{I-1}\left[\left(\frac{C_{I-j,j}}{S_j^{I+1}}\right)^2 \frac{\hat{\sigma}_j^2}{S_j^I} + \frac{\hat{\sigma}_j^2 C_{I-j,j}}{(S_j^{I+1})^2} + \hat{f}_j^2\right] - \hat{C}_{i,I}^2,$$

so

$$Var(\widehat{CDR}_i^b) = \hat{C}_{i,I}^2\left(\frac{\hat{\sigma}_{I-i}^2}{\hat{f}_{I-i}^2 S_{I-i}^I} + \frac{\hat{\sigma}_{I-i}^2}{\hat{f}_{I-i}^2 C_{i,I-i}} + 1\right) \prod_{j=I+1-i}^{I-1}\left[\left(\frac{C_{I-j,j}}{S_j^{I+1}}\right)^2 \frac{\hat{\sigma}_j^2}{\hat{f}_j^2 S_j^I} + \frac{\hat{\sigma}_j^2 C_{I-j,j}}{\hat{f}_j^2(S_j^{I+1})^2} + 1\right] - \hat{C}_{i,I}^2.$$

As $\frac{\hat{\sigma}_{I-i}^2}{C_{i,I-i}} \approx 0$, we use the linear approximation

$$\prod_j (1 + x_j) \approx 1 + \sum_j x_j,$$

and we obtain

$$Var(\widehat{CDR}_i^b) \approx \hat{C}_{i,I}^2\left(1 + \frac{\hat{\sigma}_{I-i}^2}{\hat{f}_{I-i}^2 S_{I-i}^I} + \frac{\hat{\sigma}_{I-i}^2}{\hat{f}_{I-i}^2 C_{i,I-i}} + \sum_{j=I-i+1}^{I-1}\left(\frac{C_{I-j,j}}{S_j^{I+1}}\right)^2 \frac{\hat{\sigma}_j^2}{\hat{f}_j^2 S_j^I} + \frac{\hat{\sigma}_j^2 C_{I-j,j}}{\hat{f}_j^2(S_j^{I+1})^2}\right) - \hat{C}_{i,I}^2.$$

Finally, we obtain a split of the total variance into estimation error and process error:

$$Var(\widehat{CDR}_i^b) \approx \hat{C}_{i,I}^2 \underbrace{\left( \frac{\hat{\sigma}_{I-i}^2}{\hat{f}_{I-i}^2 S_{I-i}^I} + \sum_{j=I-i+1}^{I-1} \left(\frac{C_{I-j,j}}{S_j^{I+1}}\right)^2 \frac{\hat{\sigma}_j^2}{\hat{f}_j^2 S_j^I} \right)}_{\text{estimation error}}$$
$$+ \hat{C}_{i,I}^2 \underbrace{\left( \frac{\hat{\sigma}_{I-i}^2}{\hat{f}_{I-i}^2 C_{i,I-i}} + \sum_{j=I-i+1}^{I-1} \frac{\hat{\sigma}_j^2 C_{I-j,j}}{\hat{f}_j^2 (S_j^{I+1})^2} \right)}_{\text{process error}}. \quad (13)$$

**Remark:** The process error is written as the first-order linear approximation in $\frac{\hat{\sigma}_{I-i}^2}{C_{i,I-i}}$ of equation (3). Thus, for a single accident year $i$, the prediction error is written, at first order in $\frac{\hat{\sigma}_{I-i}^2}{C_{i,I-i}}$, as the sum of the estimation error and process error, those being then orthogonal.

## Prediction error for aggregated accident years

The variance of the aggregate CDR is

$$Var\left(\sum_{i=1}^I \widehat{CDR}_i^b\right) = \sum_{i=1}^I Var(\widehat{CDR}_i^b) + 2 \sum_{1 \leq i < j \leq I} Cov(\widehat{CDR}_i^b, \widehat{CDR}_j^b),$$

with $Cov(\widehat{CDR}_i^b, \widehat{CDR}_j^b) = \mathbb{E}(\widehat{CDR}_i^b \widehat{CDR}_j^b) - \mathbb{E}(\widehat{CDR}_i^b)\mathbb{E}(\widehat{CDR}_j^b)$.

We have

$$\mathbb{E}[\widehat{CDR}_i^b]$$
$$= \mathbb{E}\left[ C_{i,I-i} \prod_{j=I-i}^{I-1} \hat{f}_j - (f_{I-i}^{b,I} C_{i,I-i} + \hat{\sigma}_{I-i}\sqrt{C_{i,I-i}}\, \epsilon_{I-i}^b) \prod_{j=I+1-i}^{I-1} \left( \frac{S_j^I}{S_j^{I+1}} \hat{f}_j + \frac{f_j^{b,I} C_{I-j,j}}{S_j^{I+1}} + \frac{\hat{\sigma}_j\sqrt{C_{I-j,j}}\epsilon_j^b}{S_j^{I+1}} \right) \right],$$

and using the independence properties previously detailed, we obtain

$$\mathbb{E}[\widehat{CDR}_i^b]$$
$$= C_{i,I-i} \prod_{j=I-i}^{I-1} \hat{f}_j - \mathbb{E}[f_{I-i}^{b,I} C_{i,I-i} + \hat{\sigma}_{I-i}\sqrt{C_{i,I-i}}\, \epsilon_{I-i}^b] \prod_{j=I+1-i}^{I-1} \mathbb{E}\left[ \frac{S_j^I}{S_j^{I+1}} \hat{f}_j + \frac{f_j^{b,I} C_{I-j,j}}{S_j^{I+1}} + \frac{\hat{\sigma}_j\sqrt{C_{I-j,j}}\epsilon_j^b}{S_j^{I+1}} \right],$$

i.e. $\mathbb{E}[\widehat{CDR}_i^b] = 0$.

So, the covariance is written

$$Cov(\widehat{CDR}_i^b, \widehat{CDR}_j^b) = \mathbb{E}(\widehat{CDR}_i^b \widehat{CDR}_j^b).$$

We have

$$Cov(\widehat{CDR}_i^b, \widehat{CDR}_j^b)$$
$$= \mathbb{E}\left[ \begin{array}{l} \left( C_{i,I-i} \prod_{k=I-i}^{I-1} \hat{f}_k - (f_{I-i}^{b,I} C_{i,I-i} + \hat{\sigma}_{I-i}\sqrt{C_{i,I-i}}\, \epsilon_{I-i}^b) \prod_{k=I+1-i}^{I-1} \left( \frac{S_k^I}{S_k^{I+1}} \hat{f}_k + \frac{f_k^{b,I} C_{I-k,k}}{S_k^{I+1}} + \frac{\hat{\sigma}_k\sqrt{C_{I-k,k}}\epsilon_k^b}{S_k^{I+1}} \right) \right) \\ \times \left( C_{j,I-j} \prod_{k=I-j}^{I-1} \hat{f}_k - \left(f_{I-j}^{b,I} C_{j,I-j} + \hat{\sigma}_{I-j}\sqrt{C_{j,I-j}}\, \epsilon_{I-j}^b\right) \prod_{k=I+1-j}^{I-1} \left( \frac{S_k^I}{S_k^{I+1}} \hat{f}_k + \frac{f_k^{b,I} C_{I-k,k}}{S_k^{I+1}} + \frac{\hat{\sigma}_k\sqrt{C_{I-k,k}}\epsilon_k^b}{S_k^{I+1}} \right) \right) \end{array} \right].$$

We suppose $i < j$ thus $I - i > I - j$ and

$$Cov(\widehat{CDR}_i^b, \widehat{CDR}_j^b)$$
$$= -\left(C_{i,I-i}\prod_{k=I-i}^{I-1}\hat{f}_k\right)\left(C_{j,I-j}\prod_{k=I-j}^{I-1}\hat{f}_k\right)$$
$$+ \underbrace{\mathbb{E}\left[\left(f_{I-i}^{b,I}C_{i,I-i} + \hat{\sigma}_{I-i}\sqrt{C_{i,I-i}}\,\epsilon_{I-i}^b\right)\left(\frac{S_{I-i}^I}{S_{I-i}^{I+1}}\hat{f}_{I-i} + \frac{f_{I-i}^{b,I}C_{i,I-i}}{S_{I-i}^{I+1}} + \frac{\hat{\sigma}_{I-i}\sqrt{C_{i,I-i}}\epsilon_{I-i}^b}{S_{I-i}^{I+1}}\right)\right]}_{\gamma_i}$$
$$\times \prod_{k=I+1-i}^{I-1}\underbrace{\mathbb{E}\left[\left(\frac{S_k^I}{S_k^{I+1}}\hat{f}_k + \frac{f_k^{b,I}C_{I-k,k}}{S_k^{I+1}} + \frac{\hat{\sigma}_k\sqrt{C_{I-k,k}}\epsilon_k^b}{S_k^{I+1}}\right)^2\right]}_{\delta_k}.$$

**Calculation of $\gamma_i$**

Using the independence properties, we have
$$\gamma_i = \frac{S_{I-i}^I}{S_{I-i}^{I+1}}C_{i,I-i}\hat{f}_{I-i}^2 + \frac{C_{i,I-i}^2}{S_{I-i}^{I+1}}\mathbb{E}\left[\left(f_{I-i}^{b,I}\right)^2\right] + \frac{\hat{\sigma}_{I-i}^2 C_{i,I-i}}{S_{I-i}^{I+1}}\mathbb{E}\left[\left(\epsilon_{I-i}^b\right)^2\right],$$
$$\gamma_i = \frac{S_{I-i}^I}{S_{I-i}^{I+1}}C_{i,I-i}\hat{f}_{I-i}^2 + \frac{C_{i,I-i}^2}{S_{I-i}^{I+1}}\left(\frac{\hat{\sigma}_{I-i}^2}{S_{I-i}^I} + \hat{f}_{I-i}^2\right) + \frac{\hat{\sigma}_{I-i}^2 C_{i,I-i}}{S_{I-i}^{I+1}},$$
$$\gamma_i = C_{i,I-i}\hat{f}_{I-i}^2 + \frac{\hat{\sigma}_{I-i}^2 C_{i,I-i}}{S_{I-i}^{I+1}} + \frac{C_{i,I-i}^2}{S_{I-i}^{I+1}}\frac{\hat{\sigma}_{I-i}^2}{S_{I-i}^I}.$$

**Calculation of $\delta_k$**

$$\delta_k = Var\left(\frac{S_k^I}{S_k^{I+1}}\hat{f}_k + \frac{f_k^{b,I}C_{I-k,k}}{S_k^{I+1}} + \frac{\hat{\sigma}_k\sqrt{C_{I-k,k}}\epsilon_k^b}{S_k^{I+1}}\right) + \mathbb{E}\left[\frac{S_k^I}{S_k^{I+1}}\hat{f}_k + \frac{f_k^{b,I}C_{I-k,k}}{S_k^{I+1}} + \frac{\hat{\sigma}_k\sqrt{C_{I-k,k}}\epsilon_k^b}{S_k^{I+1}}\right]^2.$$

Using the independence properties, we also have
$$\delta_k = \left(\frac{C_{I-k,k}}{S_k^{I+1}}\right)^2 Var(f_k^{b,I}) + \frac{\hat{\sigma}_k^2 C_{I-k,k}}{(S_k^{I+1})^2} + \hat{f}_k^2,$$

i.e.
$$\delta_k = \left(\frac{C_{I-k,k}}{S_k^{I+1}}\right)^2 \frac{\hat{\sigma}_k^2}{S_k^I} + \frac{\hat{\sigma}_k^2 C_{I-k,k}}{(S_k^{I+1})^2} + \hat{f}_k^2.$$

The covariance is thus written
$$Cov(\widehat{CDR}_i^b, \widehat{CDR}_j^b)$$
$$= \hat{C}_{i,I}\hat{C}_{j,I}\left(1 + \frac{C_{i,I-i}\hat{\sigma}_{I-i}^2}{S_{I-i}^{I+1}\hat{f}_{I-i}^2 S_{I-i}^I} + \frac{\hat{\sigma}_{I-i}^2}{\hat{f}_{I-i}^2 S_{I-i}^{I+1}}\right)\prod_{k=I+1-i}^{I-1}\left(1 + \left(\frac{C_{I-k,k}}{S_k^{I+1}}\right)^2\frac{\hat{\sigma}_k^2}{\hat{f}_k^2 S_k^I} + \frac{\hat{\sigma}_k^2 C_{I-k,k}}{\hat{f}_k^2(S_k^{I+1})^2}\right) - 1.$$

Using a linear approximation, we finally obtain

$$Cov(\widehat{CDR}_i^b, \widehat{CDR}_j^b)$$
$$= \hat{C}_{i,I}\hat{C}_{j,I}\underbrace{\left(\frac{C_{i,I-i}\hat{\sigma}_{I-i}^2}{S_{I-i}^{I+1}\hat{f}_{I-i}^2 S_{I-i}^I} + \sum_{k=I-i+1}^{I-1}\left(\frac{C_{I-k,k}}{S_k^{I+1}}\right)^2\frac{\hat{\sigma}_k^2}{\hat{f}_k^2 S_k^I}\right)}_{\text{Covariance of the estimation error}}$$
$$+ \hat{C}_{i,I}\hat{C}_{j,I}\underbrace{\left(\frac{\hat{\sigma}_{I-i}^2}{\hat{f}_{I-i}^2 S_{I-i}^{I+1}} + \sum_{k=I-i+1}^{I-1}\frac{\hat{\sigma}_k^2 C_{I-k,k}}{\hat{f}_k^2(S_k^{I+1})^2}\right)}_{\text{Covariance of the process error}}. \quad (14)$$

**Remark:** The covariance of the process error is written as a first-order linear approximation in $\frac{\widehat{\sigma}_k^2}{S_k^{I+1}}$ of equation (4).

Finally, results (13) and (14) allow to write the prediction error for aggregated accident years as the sum of the aggregated estimation error and the aggregated process error at a first-order approximation in $\frac{\widehat{\sigma}_k^2}{C_{k,I-k}}$:

$$Var\left(\sum_{i=1}^{I} \widehat{CDR}_i^b\right) = Var\left(\sum_{i=1}^{I} \widehat{CDR}_i\right) + Var\left(\sum_{i=1}^{I} CDR_i\right).$$